\documentclass{aa}  

\usepackage{graphicx}
\usepackage{txfonts}
\usepackage[english=american]{csquotes}
\usepackage{bm}
\usepackage{enumitem}
\usepackage{bbold}
\usepackage{hyperref}
\usepackage{comment}
\usepackage{xcolor}
\newcommand{\cmfast}{\textsc{{\tt{21cmFAST}}}}
\begin{document}

   \title{How informative are summaries of the cosmic 21 cm signal?}

   \author{David Prelogović
          \and
          Andrei Mesinger
          }

   \institute{Scuola Normale Superiore, Piazza dei Cavalieri 7, 56125 Pisa, Italy\\
              \email{david.prelogovic@sns.it}
    }

  \date{Received - - - / Accepted - - -}

  \abstract{
The cosmic 21 cm signal will bring data-driven advances to studies of Cosmic Dawn (CD) and the Epoch of Reionization (EoR). Radio telescopes such as the Square Kilometre Array (SKA) will eventually map the HI fluctuations over the first billion years -- the majority of our observable Universe. With such large data volumes, it becomes increasingly important to develop \enquote{optimal} summary statistics, which will allow us to learn as much as possible about the CD and EoR.
In this work we compare the astrophysical parameter constraining power of several 21 cm summary statistics, using the determinant of the Fisher information matrix,  $\det \bm{F}$.  Since we do not have an established \enquote{fiducial} model for the astrophysics of the first galaxies, we computed for each summary the distribution of $\det \bm{F}$ across the prior volume.
Using a large database of cosmic 21 cm light cones that include realizations of telescope noise, we compared the following summaries:
  (i) the spherically averaged power spectrum (1DPS), (ii) the cylindrically averaged power spectrum (2DPS), (iii) the 2D wavelet scattering transform (WST), (iv) a recurrent neural network (RNN) trained as a regressor; (v) an information-maximizing neural network (IMNN); and (vi) the combination of 2DPS and IMNN.
  Our best performing individual summary is the 2DPS, which provides relatively high Fisher information throughout the parameter space.  
  Although capable of achieving the highest Fisher information for some parameter choices, the IMNN does not generalize well, resulting in a broad distribution across the prior volume.
 Our best results are achieved with the concatenation of the 2DPS and IMNN.
The combination of only these two complimentary summaries reduces the recovered parameter variances on average by factors of $\sim$6.5 -- 9.5, compared with using each summary independently.
  Finally, we point out that that the common assumption of a constant covariance matrix when doing Fisher forecasts using 21 cm summaries can significantly underestimate parameter constraints.

  }


   \maketitle
%

\section{Introduction}
The cosmic 21 cm signal, corresponding to the spin flip transition of the ground state of HI, provides a window on the first billion years of the Universe's evolution.  This under-explored period witnessed fundamental cosmic milestones, including the Cosmic Dawn (CD) of the first galaxies and the final phase change of our Universe: the Epoch of Reionization (EoR). Current radio interferometers are setting increasingly tight upper limits on its power spectrum (PS; e.g.,{  \citealt{Trott2020, Mertens2020, Gehlot2020, HERA2023, Munshi2024}}), while the upcoming Square Kilometre Array (SKA)\footnote{\url{https://www.skatelescope.org}} promises to provide a 3D image of the EoR in the next decade(s) (e.g., \citealt{Koopmans2015, Mesinger2020}).
The unknown properties of the first galaxies and intergalactic medium (IGM)  structures are encoded in the timing and morphology of the cosmic 21 cm signal (e.g., \citealt{Pritchard2007, McQuinn2012, Visbal2012, Pacucci2014}).

The only statistically robust way to infer these properties is through Bayesian inference: sampling from theory and comparing forward models to observations. But what should we use to compare the forward model to the data?; in other words, how should we construct the likelidhood? The cosmic 21 cm signal is intrinsically a 3D light cone map.  Performing inference directly on the light cone would be incredibly challenging due to the high dimensionality of the data.
Effectively, at each frequency the SKA should obtain a sky map comparable to current cosmic microwave background maps, but with the full light cone including thousands of such frequency slices (e.g., \citealt{Loeb2004}).\footnote{Inference on high-dimensional light cones is still possible in some cases, for example galaxy large-scale structure maps (e.g., \citealt{Kitaura2008, Jasche2010, Jasche2012, Jasche2013, Leclercq2017, McAlpine2022, Dai2022, Bayer2023}).  Although such studies still compress the full galaxy observations, only treating them as biased tracers of the matter field, they are able to recover the phase information of our Universe's initial conditions (so-called constrained realizations).  It remains to be seen if such studies can be extended to the EoR and CD (though see \citealt{Zhao2023a} for a recent proof-of-concept study).}  Instead, the data are compressed into a summary statistic.
Summary statistics usually involve some form of averaging, which increases the signal-to-noise ratio (S/N)
and can motivate assuming an approximately Gaussian likelihood when performing inference (e.g., \citealt{Greig2015, Gazagnes2021, Watkinson2022}).\footnote{The Gaussianity of summaries constructed by averaging can be loosely motivated by appealing to the Central Limit Theorem. However, the theorem holds only for independent, identically distributed random variables. Unfortunately, we only have a single observable Universe; thus summaries of cosmological datasets (e.g., the PS) often involve binning over correlated (i.e., not independent) and/or differently evolving fields (i.e., not identically distributed), resulting in non-Gaussian likelihoods (e.g., \citealt{Prelogovic2023}).}

The most common choice of a 21 cm summary statistic is the spherically averaged power spectrum (1DPS).
The 1DPS maximizes the S/N of an interferometric observation -- a primary concern  in the run-up to a first detection (e.g., \citealt{Greig2018, Trott2020, Mertens2020, HERA2023}).
Another common summary choice is the cylindrically averaged power spectrum (2DPS), which includes additional information on the anisotropy of the cosmic 21 cm signal (e.g., \citealt{Bharadwaj2005b, Barkana2006, Datta2012, Mao2012}). Fourier space is also a natural basis for isolating the cosmic signal from instrumental effects and foregrounds, which primarily reside in a wedge region in the 2DPS (e.g., \citealt{Morales2012,Vedantham2012,Trott2012,Parsons2014,Liu2014a,Liu2014b,Murray2018}). However, the PS ignores Fourier phases, which can encode significant information for a highly non-Gaussian signal such as the 21 cm signal from the EoR and CD (e.g., \citealt{Bharadwaj2005a, Mellema2006, Shimabukuro2016, Majumdar2018, Watkinson2019}).  Because of this, several studies have also explored non-Gaussian statistics, such as the bispectrum \citep{Shimabukuro2016, Majumdar2018, Watkinson2019, Watkinson2022}, the trispectrum \citep{Cooray2008, Floss2022}, morphological spectra \citep{Gazagnes2021}, the bubble size distribution \citep{Lin2016, Giri2018, Shimabukuro2022, Doussot2022, Lu2024}, one-point statistics \citep{Harker2009, Watkinson2015, Kittiwisit2022}, genus topology \citep{Hong2014}, wavelet scattering transform \citep[WST;][]{Greig2022, Greig2023, Zhao2023b}, Minkowski functionals and tensors \citep{Gleser2006, Yoshiura2017, Chen2019, Kapahtia2019, Spina2021}, and Betti numbers \citep{Giri2021, Kapahtia2021}.

Is there an \enquote{optimal} choice of summary statistic?  One way to define optimal is by how tightly we can recover cosmological and astrophysical parameters.   This can be quantified through the Fisher information matrix \citep{Fisher1935}.  The Fisher information is commonly used to make forecasts using predetermined summary statistics, or even to define the summary statistic itself. Optimal summaries (algorithms) that are based on the Fisher matrix include Karhunen-Loève methods and massively optimized parameter estimation and data compression for linear transformations (e.g., \citealt{Tegmark1997, Heavens2000}), as well as nonlinear generalizations like information-maximizing neural networks (IMNNs; \citealt{Charnock2018}) and fishnets \citep{Makinen2023}. However, these  summary statistics come at the cost of interpretability, making the compression more of a \enquote{black box} compared to physically motivated and easy-to-interpret summaries like the PS.\footnote{An interesting question is why would we even expect a truly optimal (lossless) compression to exist? The answer to this can be motivated through the {manifold hypothesis} \citep{Fefferman2016}, which states that physical probability distributions of the data often lie on a low-dimensional manifold. Optimal (lossless) compression would then \enquote{extract} such manifolds from the original high-dimensional data. The above-mentioned algorithms do not attempt this directly, but aim at making compression as optimal as possible, given certain assumptions.}

It is also important to note that, unlike  $\Lambda$ cold dark matter ($\Lambda$CDM), there is no obvious or unique choice for parameterizing the astrophysics of early galaxies whose radiation determines the cosmic 21 cm signal. Nor do we have a good idea of a \enquote{fiducial} model. Thus, there is no guarantee that a summary statistic that maximizes Fisher information at a single parameter value for a single model parametrization would generalize well to other parameters and models.

Here we systematically compare the information content of several common 21 cm summaries on the basis of their Fisher information. These include: (i) the 1DPS, (ii) the 2DPS, (iii) the 2D WST, (iv) a recurrent neural network (RNN) trained as a regressor; and (v) an IMNN.\footnote{  Note that the dimensionality of the PS (1D or 2D) labels the dimensionality of the summary, while for the WST, 2D labels the dimensionality of the data on which it is computed. In our case, those are 2D sky-plane images.}
Several previous studies compared the constraining power of 21 cm summaries (e.g., \citealt{Gazagnes2021, Watkinson2022, Greig2022, Zhao2023a, Hothi2023}).  Here we explore a broader range of summary statistics, introducing IMNNs to the field. Unlike previous 21 cm Fisher studies, we do not make the simplifying assumption of a constant covariance matrix.
Most importantly, we compare the summary statistics across a broad parameter space, instead of choosing only one or two fiducial values to make a mock 21 cm observation.  This is very important for verifying the robustness of the summary since we do not have a {fiducial} astrophysical model for the first galaxies.

The paper is organized as follows. In Sect. \ref{sec:fisher_matrix} we define the information content of a summary through the Fisher matrix and introduce the Fisher information distribution. In Sect. \ref{sec:21cm_and_summaries} we introduce our 21 cm simulator and explain the summaries we consider in detail. In Sect.\ \ref{sec:database} we discuss our database of prior samples for evaluating the Fisher information. In Sect. \ref{sec:results} we present our results, quantifying the most informative summaries as well as the error introduced by the common assumption of a constant covariance matrix.  Finally, we present  concluding remarks and future prospects in Sect. \ref{sec:conclusions}. All quantities are quoted in comoving units, and we assume a standard $\Lambda\mathrm{CDM}$ cosmology: $\left(\Omega_{\Lambda}, \Omega_{\mathrm{M}}, \Omega_{b}, n, \sigma_{8}, H_{0}\right) =\left(0.69,0.31,0.048,0.97,0.81,68 \mathrm{~km} \mathrm{~s}^{-1} \mathrm{Mpc}^{-1}\right)$, consistent with the results from \cite{Planck2020VI}.  

\section{The Fisher matrix as a measure of information content} \label{sec:fisher_matrix}
How do we measure the information content of a summary?  If we have a theoretical model with a given parametrization, we can see how sensitive the summary is to changes in the astrophysical/cosmological parameters around some fiducial value. 
 Specifically, we can calculate the Fisher matrix (e.g., \citealt{Spall2005}):
\begin{align}
    \bm{F}(\bm{\theta}^*)_{mn} &= \int \mathrm{d}\bm{d} \, P(\bm{d} | \bm{\theta}^*)\,\,\, \frac{\partial}{\partial\bm{\theta}_m} \ln P(\bm{d} | \bm{\theta}^*) \cdot \frac{\partial}{\partial\bm{\theta}_n} \ln P(\bm{d} | \bm{\theta}^*) \nonumber\\
    &= \mathbb{E}\left[\left.\frac{\partial}{\partial\bm{\theta}_m} \ln P(\bm{d} | \bm{\theta}) \cdot \frac{\partial}{\partial\bm{\theta}_n} \ln P(\bm{d} | \bm{\theta})\right|{\bm{\theta^*}} \right] \, ,
    \label{eq:Fisher_def}
\end{align}
where $\bm{d}$ is the data summary (c.f. Fig. \ref{fig:pipeline}), $P(\bm{d} | \bm{\theta})$ is the likelihood, and $\mathbb{E}$ denotes the expectation value over the likelihood (i.e., an empirical average over many realizations) evaluated at a given point in parameter space $\bm{\theta}^*$. The Fisher information matrix provides the {maximum} constraining power, known as the Cram\'er-Rao bound \citep{Rao1992, Cramer1999}: $\operatorname{Var}(\hat{\bm{\theta}}_m) \ge (\bm{F}^{-1})_{mm}$, 
where $\hat{\bm{\theta}}$ is an unbiased estimator of the parameters.  
In the multivariate case, one can equivalently write:
\begin{equation}
    \operatorname{Cov}(\hat{\bm{\theta}}) \ge \bm{F}^{-1} \, ,
\end{equation}
where the matrix inequality is interpreted as $\operatorname{Cov}(\hat{\bm{\theta}}) - \bm{F}^{-1}$ being a positive semi-definite matrix. One can then prove the following inequality (see Appendix \ref{app:det_proof}):
\begin{equation}
    \det{\operatorname{Cov}(\hat{\bm{\theta}})} \ge \det{\bm{F}^{-1}} \, .
\end{equation}
Therefore, the determinant of the Fisher matrix measures the tightest parameter volume that can be constrained by the data.  From now on, {we refer to  $\det \bm{F}(\bm{\theta}^*)$ as simply the  Fisher information around $\bm{\theta}^*$} --- the higher its value, the more constraining the summary is around $\bm{\theta}^*$.

\subsection{Gaussian approximation}

In almost all cosmological scenarios, we do not explicitly know the full likelihood function $P(\bm{d} | \bm{\theta})$, nor its \enquote{score,} $\nabla_{\bm{\theta}} \ln P(\bm{d} | \bm{\theta})$. For performing quick forecasts, it is instead common to approximate the likelihood as a Gaussian {in data space} $P(\bm{d} | \bm{\theta}) \approx \mathcal{N}(\bm{d} | \bm{\mu}(\bm{\theta}), \bm{\Sigma}(\bm{\theta}))$, where the mean $\bm{\mu}(\bm{\theta})$ and covariance $\bm{\Sigma}(\bm{\theta})$ are general functions of the parameter space.

Assuming a Gaussian likelihood, one can compute the integral in Eq. \ref{eq:Fisher_def} analytically (Appendix \ref{app:gaussian_fisher}; see also \citealt{Tegmark1997, Vogeley1996}):
\begin{align}
    \bm{F}_{mn} &\equiv \bm{F}_{\bm{\mu}, \, mn} + \bm{F}_{\bm{\Sigma}, \, mn} \nonumber\\
    &= \frac{\partial \bm{\mu}^T}{\partial\theta_m} \bm{\Sigma}^{-1} \frac{\partial \bm{\mu}}{\partial\theta_n} + \frac{1}{2} \operatorname{tr} \left[\bm{\Sigma}^{-1} \frac{\partial \bm{\Sigma}}{\partial \theta_m} \bm{\Sigma}^{-1} \frac{\partial \bm{\Sigma}}{\partial \theta_n} \right]\, . \label{eq:Fisher_Gaussian}
\end{align}
Here the first (second) term measures how the mean (covariance) of the data summary changes with respect to the parameters.
Cosmological Fisher forecasts usually ignore one of the two terms (most often the second term $\bm{F}_{\bm{\Sigma}}$), effectively fixing either the mean or the covariance matrix to some fiducial value. However, we will see below that both terms are non-negligible for some summaries (e.g., \citealt{Carron2013}).

\subsection{Finite differencing}
To numerically evaluate Eq. \ref{eq:Fisher_Gaussian}, we grouped the terms into \enquote{fixed} ($\bm{\mu}$, $\bm{\Sigma}$, $\bm{\Sigma}^{-1}$) and \enquote{differentiated} ($\nabla_{\bm{\theta}} \, \bm{\mu}$, $\nabla_{\bm{\theta}} \, \bm{\Sigma}$). We calculated {fixed} terms from realizations of the summary around the same parameter value $\bm{\theta}^*$. To compute the mean, we can use the maximum likelihood estimator (MLE):
\begin{equation}
    \bm{\mu}(\bm{\theta}^*) = \frac{1}{N} \sum_{i=1}^N \bm{d}(\bm{\theta}^*)_i \, , \label{eq:mu_mean}
\end{equation}
where the index $i$ labels different realizations of initial conditions and other sources of stochasticity (e.g., telescope noise). 
It is known that such an estimator of the mean is unbiased. However, for the covariance matrix the situation is more complex, as the MLE
\begin{equation}
    \bm{\Sigma}_{\text{MLE}}(\bm{\theta}^*) = \frac{1}{N} \sum_{i=1}^N (\bm{d}(\bm{\theta}^*)_i - \bm{\mu}(\bm{\theta^*})) (\bm{d}(\bm{\theta}^*)_i - \bm{\mu}(\bm{\theta}^*))^T \,  \label{eq:sigma_mean}
\end{equation}
is biased. One can show that the estimators of the covariance and its inverse can be unbiased in the following way (e.g., \citealt{Hartlap2007}):
\begin{align}
    \bm{\Sigma}(\bm{\theta}^*) &= \frac{N}{N - 1} \, \bm{\Sigma}_{\text{MLE}}(\bm{\theta}^*) \, , \label{eq:sigma_unbiased}\\
    \bm{\Sigma}^{-1}(\bm{\theta}^*) &= \frac{N - D - 2}{N - 1} \, \bm{\Sigma}_{\text{MLE}}^{-1}(\bm{\theta}^*)  \, , \label{eq:sigma_inv_unbiased}
\end{align}
where $D$ is dimensionality of the summary $\bm{d}$. The correction prefactor for the inverse of the covariance is particularly important when $D$ is close to the number of samples $N$. Without it, both $\bm{F}_{\bm{\mu}}$ and $\bm{F}_{\bm{\Sigma}}$ would end up over-confident, with the effect on $\bm{F}_{\bm{\Sigma}}$ being quadratic (see Eq. \ref{eq:Fisher_Gaussian}).

A similar approach is used to estimate the \enquote{differentiated parameters.} In the case in which the simulator is not differentiable, we can use finite differencing to estimate gradients of the mean and covariance:
\begin{align}
    \frac{\delta\bm{d}(\bm{\theta}^*)_i}{\delta\bm{\theta}_m} &\equiv \frac{\bm{d}(\bm{\theta}^* + \delta\bm{\theta}_m / 2)_i - \bm{d}(\bm{\theta}^* - \delta\bm{\theta}_m / 2)_i}{\delta\bm{\theta}_m} \, , \label{eq:numerical_derivative_sample}\\
    \left.\frac{\partial\bm{\mu}}{\partial\bm{\theta}_m}\right|_{\bm{\theta}^*} &= \frac{1}{M}\sum_{i=1}^M \, \frac{\delta\bm{d}(\bm{\theta}^*)_i}{\delta\bm{\theta}_m} \, , \label{eq:mu_der}\\
    \left.\frac{\partial\bm{\Sigma}}{\partial\bm{\theta}_m}\right|_{\bm{\theta}^*} &= \frac{2}{L - 1}\sum_{i=1}^{L} \left(\frac{\delta\bm{d}(\bm{\theta}^*)_i}{\delta\bm{\theta}_m} - \left.\frac{\partial\bm{\mu}}{\partial\bm{\theta}_m}\right|_{\bm{\theta}^*}\right) \left(\bm{d}(\bm{\theta}^*)_i - \bm{\mu}(\bm{\theta}^*)\right)^T \, . \label{eq:sigma_der}
\end{align}
Here, $\delta\bm{\theta}_m$ represents a small change in the parameter over which the derivative is computed and $\delta\bm{d}(\bm{\theta}^*)_i / \delta\bm{\theta}_m$ corresponds to the numerical derivative of a given summary {realization} (i.e., fixing all sources of stochasticity).

\subsection{The distribution of the Fisher information over parameter space}

Below we show examples of common 21 cm summaries evaluated at a {fiducial} set of parameters, $\bm{\theta}_{\text{fid}}$.  Unfortunately, the cosmic 21 cm signal does not yet have a well-defined fiducial model and our choice of $\bm{\theta}_{\text{fid}}$ is only a \enquote{best guess.}  Therefore, to properly assess the quality of a particular summary, we sampled many points $\bm{\theta}^*$ from the prior $P(\bm{\theta})$, calculating the Fisher information $\det\bm{F}(\bm{\theta}^*)$ at each sample.  We used this {prior-weighted distribution of Fisher information as the main metric for assessing the quality of a given summary}.\footnote{  Note that the prior-weighted distribution of the Fisher information is closely related to the Mutual Information $I(\bm{x}, \bm{\theta}) = D_{\text KL} (p(\bm{x}, \bm{\theta}) \| p(\bm{x}) p(\bm{\theta}))$, where $D_{\text KL}$ is Kullback-Liebler divergence \citep{Brunel1998}. Mutual Information can also be used to asses the quality of a summary (e.g., \citealt{Sui2023}).}

\section{The cosmic 21 cm signal and its summary statistics} \label{sec:21cm_and_summaries}
\begin{figure}
    \centering
    \includegraphics[width=\linewidth]{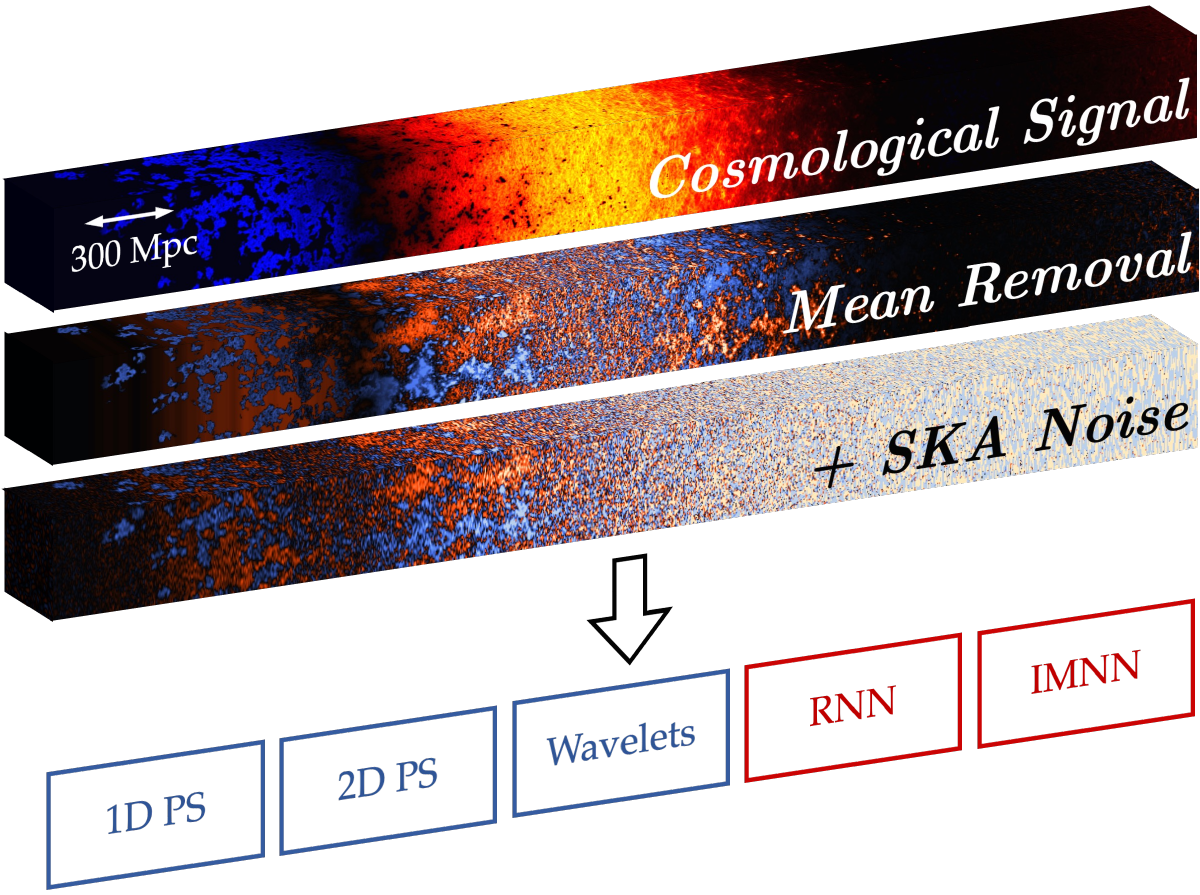}
    \caption{Schematic illustrating our pipeline for producing data summaries for a given astrophysical parameter combination (adapted from \citealt{Prelogovic2023}). Starting from a realization of the cosmological signal simulated with \cmfast{}, we remove the mean at each redshift and add a telescope noise realization corresponding to a 1000h observation with SKA1-Low. The resulting light cone is compressed into different summaries. Explicitly defined summaries are marked in blue, while neural network summaries are marked in red.  This realization is computed at the fiducial parameter set, which was used only for IMNN training and visualizations of the summaries.  See the main text for more details.}
    \label{fig:pipeline}
\end{figure}

For a given sample of astrophysical parameters, $\bm{\theta}^*$, we computed a realization of mock data using the following steps (illustrated in Fig. \ref{fig:pipeline}):
\begin{enumerate}
    \item Sampling cosmological initial conditions, we simulated a realization of the 21 cm lightcone using the \cmfast{} code.
    \item We subtracted the mean from each frequency slice, mimicking an interferometric observation.
    \item We simulated a $1000 \mathrm{h}$ tracked scan with the SKA1-low telescope, including the corresponding $uv$ coverage and realization of the instrument noise.
    \item We compressed the resulting light cone into a given summary observation.
\end{enumerate}
We elaborate on these steps below.

The cosmic 21 cm signal was simulated using the semi-numerical code, \texttt{\cmfast{} v3}\footnote{\url{https://github.com/21cmfast/21cmFAST}} \citep{Mesinger2007, Mesinger2011, 21cmFAST_JOSS}. In brief, we used the \citet{Park2019} galaxy model, varying five parameters that characterize the unknown UV and X-ray properties of high-$z$ galaxies through scaling relations:
\begin{itemize}
    \item $\bm{f_{*, 10}}$ -- the fraction of galactic gas in stars, normalized at the halo mass of $10^{10} M_\odot$.  The stellar to halo mass relation of the faint galaxies that drive reionization is well described by a power law: $M_*/M_h = f_{*, 10} (M_h/10^{10} M_\odot)^{0.5}(\Omega_b/\Omega_m)$ (see \citealt{Park2019} and references therein).
    \item $\bm{f_{\text{esc}, 10}}$ -- the ionizing UV escape fraction, normalized at the halo mass $10^{10} M_\odot$.
    Again, the (mean) escape fraction is assumed to be a power law with halo mass, and here we fixed its index to $\alpha_{\text{esc}} = -0.5$ (e.g., Qin et al. in prep).
    \item $\bm{M_{\text{turn}}}$ -- the characteristic host halo mass below which galaxies are inefficient at forming stars, due to slow gas accretion, supernovae feedback, and/or photo-heating.  
    \item $\bm{E_0}$ -- the characteristic X-ray energy below which photons are absorbed by the interstellar medium (ISM) of the host galaxy. 
    \item $\bm{L}_{\bm{X} < 2\mathrm{keV}} / \mathrm{SFR}$ -- the X-ray soft band (in the energy range $E_0$ -- 2 keV) luminosity per star formation rate (SFR), in units of $\mathrm{erg \, s^{-1} \, keV^{-1} \, M_{\odot}^{-1} \, yr}$.
    The SFR is taken to be: $\dot{M}_* = M_*/(0.5 H(z)^{-1})$, where the Hubble time, $H(z)^{-1}$, also scales with the dynamical time at the virial radius.
\end{itemize}
Our fiducial, $\bm{\theta}_{\text{fid}}$ parameter vector corresponds to the following: $\log_{10} f_{*, 10} = -1.3$, $\log_{10} f_{\text{esc}} = -1$, $\log_{10} M_{\text{turn}} [M_{\odot}] = 8.7$, $\log_{10} L_{X<2\mathrm{keV}} / \text{SFR} [\mathrm{erg \, s^{-1} \, keV^{-1} \, M_{\odot}^{-1} \, yr}] = 40$, $E_0 [\mathrm{keV}] = 0.5$. This particular choice is only used when training the IMNN, and for the visualizations of the summaries below. More details on the model and its motivations are provided in \citealt{Park2019} and references therein.

To simulate the observation, we followed the procedure described in {  \citet{Prelogovic2022, Prelogovic2023}} (c.f. Fig. \ref{fig:pipeline}). In brief, we calculated the 21 cm signal from redshifts 30 to 5, on a 300 Mpc coeval cube, with 1.5 Mpc resolution. The snapshots of the evolved coeval cube were then stacked to produce the final light cone of the signal. After the subtraction of the mean for every sky-plane slice, we added thermal noise corresponding to a a $6 \, \mathrm{h/day}$, $10 \, \mathrm{s}$ integration time tracked scan with SKA1-Low {  antenna configuration} for a  $1000 \, \mathrm{h}$ observation in total. {  For this task we used \texttt{tools21cm}\footnote{\url{https://github.com/sambit-giri/tools21cm}}, where a separate UV coverage and noise is calculated for each sky-plane slice.} Before computing each summary, we additionally smoothed the light cone with a box-car filter obtaining a final resolution of 6 Mpc. This final step does not impact our results as it is comparable to the SKA1-Low beam, but is needed in order to minimize GPU memory usage for certain summaries (see below for more details). We also note that in this work we made the optimistic assumption of perfect foreground removal, thus exploring the maximum future potential of our set of summary statistics.

The resulting 3D light cone was compressed into a given summary statistic.   We  explore the following summaries in this work:
\begin{itemize}[leftmargin=2.4cm]
    \item[1DPS] --  spherically averaged 1D power spectrum,
    \item[2DPS] --  cylindrically averaged 2D power spectrum,
    \item[Wavelets] --  2D wavelet scattering transform,
    \item[RNN] --  recurrent neural network,
    \item[IMNN] --  information-maximizing neural network.
    \item[2DPS + IMNN] --  concatenation of the 2DPS and IMNN summaries.
\end{itemize}
The first four are explicitly defined while the following two are \enquote{learned} by neural networks (NNs). The last, 2DPS + IMNN, is a combination of the two individual summaries. 
We describe them in detail below.

\subsection{Spherically averaged (1D) power spectrum}
\begin{figure}
    \centering
    \includegraphics[width=\linewidth]{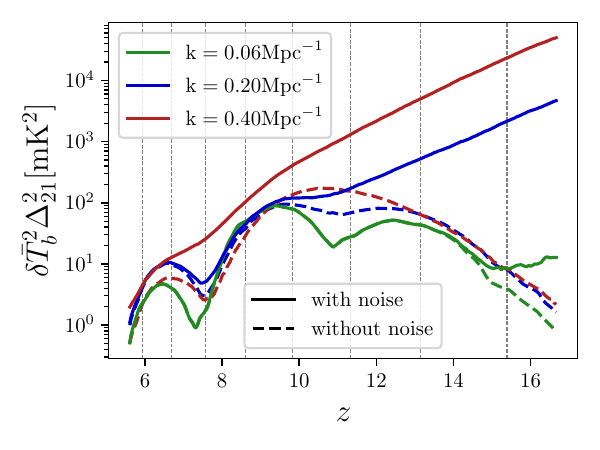}
    \caption{1DPS for the fiducial light cone shown in Figure \ref{fig:pipeline}.  
    Different colors correspond to different wave modes, while solid (dashed) lines correspond to light cones with (without) noise.  Comparing the solid and dashed lines, we see that the signal is noise dominated at higher $k$ modes and redshifts (see also Fig. \ref{fig:pipeline}).  All power spectra are computed from comoving cubes extracted from the light cone, centered on the redshifts indicated by the vertical dashed lines.}
    \label{fig:1DPS_fiducial}
\end{figure}

As discussed in the introduction, the 1DPS, is the most common choice of a summary statistic.  Despite the cosmic 21 cm signal being non-Gaussian, the 1DPS is a natural choice when seeking to maximize the S/N of an interferometric first detection.

The 1DPS of the 21 cm signal $\delta T_b(\bm{x}, z)$, where $\bm{x}$ is the sky-plane coordinate and $z$ the redshift, is defined as
\begin{equation}
    \delta\bar{T}^2_b \Delta_{21}^2(k, z') \equiv \frac{k^3}{2 \pi^2 V} \left\langle\left|\delta T_b(\bm{k}, z') - \delta\bar{T}_b(z')\right|^2\right\rangle_k \, .
    \label{eq:1DPS}
\end{equation}
The $\delta T_b(\bm{k}, z')$ is calculated from the Fourier transforms of mean-subtracted light cone segments centered around $z'$. 
Here we calculated the 1DPS in three log-spaced $k$-bins, computed at eight different $z'$.  In Fig. \ref{fig:1DPS_fiducial} we show the 1DPS of the fiducial light cone (see Fig. \ref{fig:pipeline}). Solid curves correspond to the full signal: cosmic + noise, while the dashed curves are only the cosmic signal.  We see the usual three peaked redshift evolution of the large-scale power at $k\sim$1 Mpc$^{-1}$, tracing fluctuations in the IGM ionized fraction, the IGM temperature and Lyman alpha background (e.g., \citealt{Pritchard2007, Mesinger2014}).  By comparing the solid and dashed curves, we see that the signal is noise dominated at small scales and early times.  Indeed, $k\geq$ 0.4 Mpc$^{-1}$ remains noise dominated virtually at all redshifts ($z>7$), for this fiducial model and choice of 1000h integration time.

\subsection{Cylindrical (2D) power spectrum}
\begin{figure*}
    \centering
    \includegraphics[width=12cm]{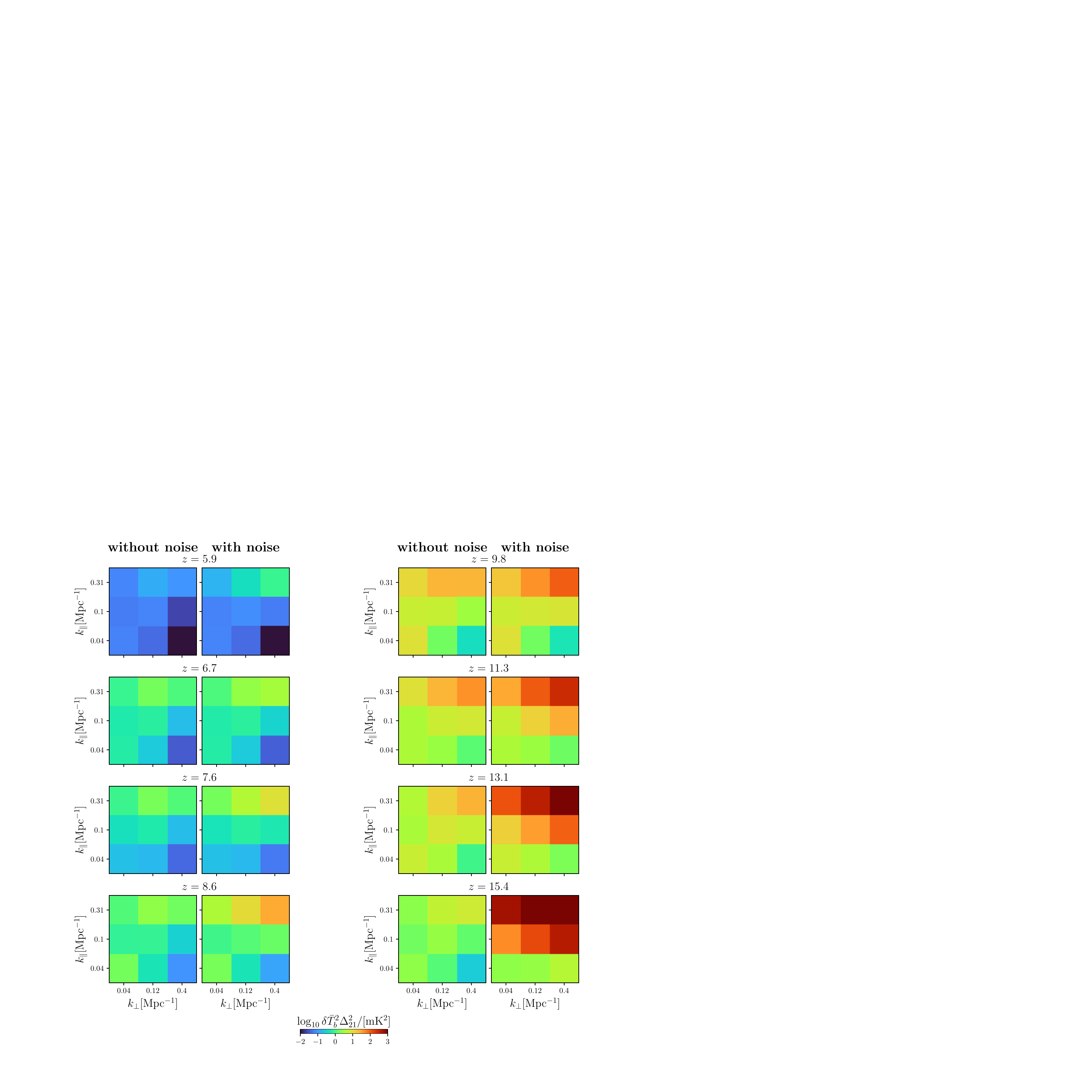}
    \caption{2DPS of the fiducial model shown in Figure \ref{fig:pipeline}. Different columns show the 2DPS for the pure cosmic signal and including thermal noise, as labeled on the top. Each pair corresponds to a different central redshift, coinciding with those used in the 1DPS summary (see Fig. \ref{fig:1DPS_fiducial}). Furthermore, the bins between 1DPS and 2DPS coincide as much as possible. As in the 1DPS case, high-$k$ modes and high redshifts are noise dominated.}
    \label{fig:2DPS_fiducial}
\end{figure*}

The cylindrically averaged 2DPS distinguishes between sky-plane ($k_\perp$) and line-of-sight ($k_\|$) modes. Redshift space distortions (e.g., \citealt{Bharadwaj2005b, Barkana2005}) and light cone evolution (e.g., \citealt{Greig2018, Mondal2018}) result in an anisotropic cosmic signal.  Therefore, the 2DPS should encode more physical information compared to the 1DPS.  Moreover, the instrument/foregrounds are better characterized in the 2DPS (e.g., see the review in \citealt{Liu2014b}), making it a natural summary observable for preliminary, low S/N measurements.

The 2DPS is defined as
\begin{equation}
    \delta\bar{T}^2_b \Delta_{21}^2(k_\perp, k_\|, z') \equiv \frac{k^2_\perp \, k_\|}{4 \pi^2 V} \left\langle\left|\delta T_b(\bm{k}, z') - \delta\bar{T}_b(z')\right|^2\right\rangle_{k_\perp, k_{\|}} \, ,
    \label{eq:2DPS}
\end{equation}
where the Fourier transform is performed on the same light cone chunks as in the 1D case, with the difference being that the expectation value is calculated over $(k_\perp, k_{\|})$ bins. Here we picked $3\times3$ log-spaced bins, aligned to the 1D case as close as possible. The 2DPS for the fiducial light cone (see Fig. \ref{fig:pipeline}) is shown in Fig. \ref{fig:2DPS_fiducial}. The panels correspond to the mean redshifts of the light cone chunks, coinciding with the ones of the 1DPS (see Fig. \ref{fig:1DPS_fiducial}). The columns correspond to the cases with and without noise, as labeled at the top.
We can verify that both the cosmic signal and the noise are anisotropic. For the cosmic signal, this is most evident for $z\lesssim10$ where the noise is subdominant. For the noise, the same can be clearly seen for higher redshifts, where the noise dominates the signal at high $k_\|$ much more than in high $k_\perp$. 

\subsection{Wavelet scattering transform}
The wavelet transform \citep{Gabor1946, Goupillaud1984, Trott2012} and WST \citep{Mallat2011} are based on convolutional filters designed on a Fourier basis with an additional Gaussian envelope along the frequency direction. As such, they capture local features in both image and Fourier domains, which is often relevant in audio and image processing. Higher-order convolutions combined with NNs (wavelet-based convolutional NNs) have been very successful for such purposes \citep{Bruna2012, Sifre2013, Anden2014}. The WST has been extensively explored in cosmological (e.g., \citealt{Cheng2020}) and 21 cm (e.g., \citealt{Greig2022, Greig2023, Zhao2023b, Hothi2023}) analyses.

Here we briefly introduce how the WST is defined. It represents a convolution of an input image $I(\bm{x})$ by a set of wavelet (here specifically Morlet) filters. They are characterized by the physical scale $j$ and rotation moment $l$. The actual physical scale of the filter $\psi^{j, l}$ will then be $2^j$ pixels, and its orientation angle $l \cdot \pi / L$. The wavelet coefficients correspond to the average of the convolved images, to which we refer as simply wavelets. We note that in this work we restricted ourselves to the 2D WST (see, e.g., \citealt{Zhao2023b} for recent work using the 3D transform).

The first- and second-order wavelets can be written as
\begin{align}
    s_1^{j_1, l_1} &= \langle|I(\bm{x}) * \psi^{j_1, l_1}|\rangle \, , \label{eq:s1_wavelets} \\
    s_2^{j_1, l_1, j_2, l_2} &= \langle||I(\bm{x}) * \psi^{j_1, l_1}| * \psi^{j_2, l_2}|\rangle \, .
\end{align}
One can see the similarity between first-order wavelets and the PS by writing $\mathrm{PS}(\bm{k}) = \langle |I(\bm{x}) * \psi'|^2 \rangle$, with $\psi' = e^{-i \bm{k}\cdot\bm{x}}$. 
To extract only cylindrically averaged information imprinted in the 21 cm signal (analogous to calculating 2DPS in a Fourier basis), one can average coefficients over the rotations:
\begin{align}
    S_1^{j_1} &= \langle s_1^{j_1, l_1}\rangle_{l_1} \, , \label{eq:S1_wavelets} \\
    S_2^{j_2} &= \langle s_2^{j_1, l_1, j_2, l_2}\rangle_{l_1, l_2} \, .
\end{align}
{  It is important to mention that this procedure removes anisotropic information. Alternative definitions such as the reduced WST \citep{Allys2019} are able to reduce the coefficient dimensionality, while keeping anisotropy.} For further information about the WST and wavelet coefficients we refer the reader to \cite{Cheng2020} and \cite{Greig2022, Greig2023}.

Our setup for calculating wavelets is the following. We first selected a series of eight images, centered at the same redshifts as for the power spectra. For each image we calculated the first- and second-order wavelets and combined them for the final summary. As the WST scale ($2^J \times 2^J$) has to be smaller than the sky-plane images ($50 \times 50$), we set $J = 4$. Furthermore, we fixed $L = 4$ \citep{Cheng2020, Greig2022}. To compute convolutions, we used the \texttt{scattering\_transform}\footnote{\url{https://github.com/SihaoCheng/scattering_transform}} package \citep{Cheng2020}.  For each redshift slice, we averaged the first- and second-order coefficients computed on a rolling 32x32 window.  Our final WST summary thus consists of $J = 4$ $S_1$ and $J (J+1) / 2 = 10$ $S_2$ coefficients for each of the eight redshift slices, for a total of 112 numbers.

\begin{figure}
    \centering
    \includegraphics[width=\linewidth]{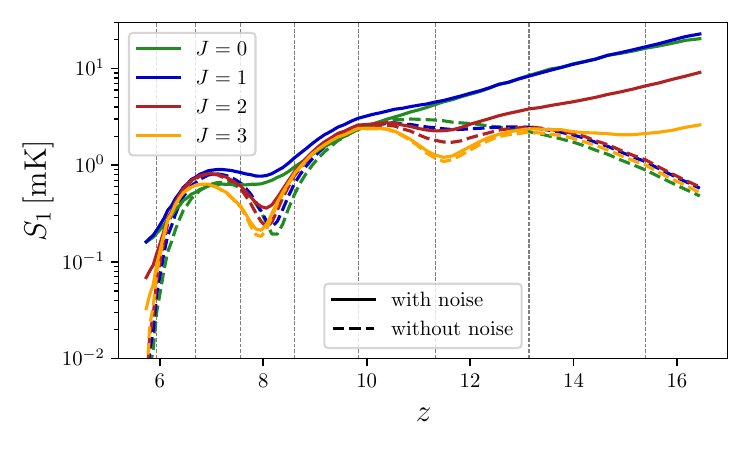}
    \caption{  First-order wavelet coefficients (Eqs. \ref{eq:s1_wavelets} and \ref{eq:S1_wavelets}) for the fiducial light cone shown in Figure \ref{fig:pipeline}.  
    Different colors correspond to different $J$ modes, while solid (dashed) lines correspond to the case with (without) noise. Comparing the solid and dashed lines, we see that the signal is noise dominated at smaller scales (smaller $J$) and higher redshifts (see also Fig. \ref{fig:pipeline}).}
    \label{fig:wavelets_fiducial}
\end{figure}
{  In Figure \ref{fig:wavelets_fiducial}, we show the redshift evolution of first-order wavelet coefficients ($S_1$; see Eqs. \ref{eq:s1_wavelets}, \ref{eq:S1_wavelets}). Solid (dashed) lines correspond to coefficients with (without) telescope noise. We see the same qualitative trends as in Fig. \ref{fig:1DPS_fiducial}, with lower $J$ modes (i.e., smaller scales) more noise dominated. As a result, these low $J$ modes contribute less to the total Fisher information \citep{Hothi2023}.}

\subsection{Recurrent neural network}
A RNN is a type of NN architecture designed to efficiently encode sequential or time-evolving information. In our previous work \citep{Prelogovic2022}, we used RNNs along the redshift axis in combination with convolutional neural networks (CNNs) along the sky-plane, to capture the anisotropic evolution of the cosmic 21 cm signal along the light cone. Specifically, we used a long short-term memory (\citealt{Hochreiter1997, Shi2015}) RNN, which is more stable to train compared with older RNN versions (see, e.g., the review in \citealt{Schmidt2019}).
Using RNNs resulted in a more accurate regression of astrophysical parameters, compared to only using CNNs (e.g., \citealt{Gillet2019, LaPlante2019, Mangena2020, Kwon2020, Zhao2022a, Heneka2023}).

In this work, we used the flagship model from \cite{Prelogovic2022}, \textit{SummaryRNN}, together with the exact weights from that work. The model has been trained in a supervised manner as a regressor. The database used for training had the same telescope simulator, but used an older version of the {\tt 21cmFAST} simulator with a different astrophysical parametrization. 

Our RNN summary in this work is the 32-dim output from the first dense layer of the network (see Table B3 in \citealt{Prelogovic2022}, where convolutional and recurrent layers are followed by dense layers, ending with prediction of four parameters). Using such a higher-dimensional layer instead of the final parameter regressor output (e.g., \citealt{Zhao2022a}), makes the summary more sensitive to the general light cone features instead of the specific model and/or parametrization used to create the training set.  This is similar to so-called transfer learning (e.g., \citealt{Jiang2022}), where one uses a NN that has been pretrained on a different domain (i.e., database) and retrains its last layers on a new domain. As our original RNN takes as input $25\times25$ sky-plane slices, we rolled a  $25\times25$ window over each redshift slice, averaging the corresponding dense layer outputs to obtain the final RNN summary.

\subsection{Information-maximizing neural network} \label{subsec:IMNN}
An IMNN (\citealt{Charnock2018})\footnote{\url{https://github.com/tomcharnock/IMNN}} is an unsupervised learning algorithm, where the NN is specifically trained to maximize the Fisher information $\det \bm{F}$ of the summary compression, at a single fixed point in parameter space  $\bm{\theta}_{\text{fid}}$.
IMNNs have been previously applied to galaxy large-scale structure maps (e.g., \citealt{Makinen2022}) and
catalog-free modeling of galaxy types \citep{Livet2021}. Here we applied them for the first time to 21 cm maps of the EoR and CD. In what follows, we outline the main ingredients of the algorithm and the training procedure.

Taking some NN architecture (in our case a simple CNN), one compresses the light cone, $\bm{l}$, into a summary vector, $\bm{d}$,
\begin{equation}
    \bm{d} = \mathrm{NN}(\bm{l}) \, .
\end{equation}
The dimensionality of the summary is equal to the parameter space dimensionality ($D$).  This is motivated by the fact that, for a given model parametrization,
the \enquote{score} of the real likelihood function, $\nabla_{\bm{\theta}} \ln P(\bm{l} | \bm{\theta})$, is sufficient for optimal parameter recovery, and thus the dimensionality of the summary that maximizes information around one fiducial point can be the same as the dimensionality of the parameter space (e.g., \citealt{Heavens2000, Charnock2018}). 

We trained the IMNN using the following loss function (c.f. \citealt{Charnock2018, Makinen2021}):

\begin{align}
    \bm{F}_{\text{IMNN}}(\bm{\theta}_{\text{fid}})_{mn} &\equiv  \left.\frac{\partial \bm{\mu}^T}{ \partial\bm{\theta}_m}\right|_{\bm{\theta}_{\text{fid}}} \, \left.\frac{\partial \bm{\mu}}{ \partial\bm{\theta}_n}\right|_{\bm{\theta}_{\text{fid}}} \label{eq:IMNN_Fisher}\\
    L_{\mathcal{F}}({\bm{\Sigma}(\bm{\theta}_{\text{fid}})}) &\equiv \| \bm{\Sigma}(\bm{\theta}_{\text{fid}})  - \mathbb{1}\|_{\mathcal{F}} \label{eq:IMNN_covariance_norm}\\
    L_2(W_{\text{NN}}) &\equiv \| W_{\text{NN}}\|^2 \label{eq:IMNN_weights_norm}\\
    \mathcal{L} &= -\ln\det \bm{F}_{\text{IMNN}} \nonumber \\
    &+\lambda \, L_{\mathcal{F}}({\bm{\Sigma}}) \tanh L_{\mathcal{F}}(\bm{\Sigma}) \label{eq:IMNN_loss}\\
    &+\lambda_W \, L_2(W_{\text{NN}}), \nonumber
\end{align}
where
\begin{align}
    \bm{\mu}(\bm{\theta_{\text{fid}}}) &= \frac{1}{N}\sum_{i = 1}^M \bm{d}(\bm{\theta}_{\text{fid}})_i \, , \label{eq:IMNN_mu}\\
    \bm{\Sigma}(\bm{\theta}_{\text{fid}}) &= \frac{1}{N - 1} \sum_{i=1}^N \left(\bm{d}(\bm{\theta}_{\text{fid}})_i - \bm{\mu}(\bm{\theta_{\text{fid}}})\right) \left(\bm{d}(\bm{\theta}_{\text{fid}})_i - \bm{\mu}(\bm{\theta}_{\text{fid}})\right)^T , \label{eq:IMNN_sigma}\\
    \left.\frac{\partial\bm{\mu}}{\partial\bm{\theta}_m}\right|_{\bm{\theta}_{\text{fid}}} &= \frac{1}{M}\sum_{i=1}^M \, \frac{\bm{d}(\bm{\theta}_{\text{fid}} + \delta\bm{\theta}_m / 2)_i - \bm{d}(\bm{\theta}_{\text{fid}} - \delta\bm{\theta}_m / 2)_i}{\delta\bm{\theta}_m} \, , \label{eq:IMNN_dmu}
\end{align}
$L_{\mathcal{F}}({\bm{\Sigma}(\bm{\theta}_{\text{fid}})})$ is the Frobenius norm between the covariance and the unit matrix $\bm{\mathbb{1}}$, $L_2(W_{NN})$ is the $l_2$ norm of the networks' weights. The final loss is then the sum of the negative logarithm of the Fisher information, together with the covariance and weight regularizers. The $\tanh$ term in the covariance regularizer is used to turn-off the regularization once the covariance becomes close to the unity. 

The main goal of the IMNN is therefore to maximize the Fisher information, while keeping the covariance of the summary fixed to unity. {  The reason we have the flexibility to normalize the covariance of the summary is the following. For any summary $\bm{d}$ of dimensionality D, one can define a new summary, $\bm{d}' = \bm{a} + \bm{B} \bm{d}$, where $\bm{a}$ is a constant D-dim vector serving to normalize $\bm{d}'$ to be zero mean, and $\bm{B}$ is a nonsingular $D\times D$ matrix, serving to normalize $\bm{d}'$ to be unit variance ($\bm{B}$ is equivalent to a zero-phase component analysis (ZCA) whitening kernel; see \citealt{Kessy2018}). One can then show that $\ln P(\bm{d} | \bm{\theta}) = \ln P(\bm{d}' | \bm{\theta}) + \ln \det \bm{B} \,$, or equivalently $\partial \ln P(\bm{d} | \bm{\theta}) / \partial\bm{\theta}_m = \partial\ln P(\bm{d}' | \bm{\theta}) / \partial\bm{\theta}_m$. This means there is a freedom to transform the summary $\bm{d}$ to be zero-mean, unit-variance {at a single point} $\bm{\theta_{\rm fid}}$, without changing the Fisher information anywhere in the parameter space.} 

{  We also note that the Fisher matrix used for computing the IMNN loss is calculated with two approximations: i) using only the first term; and ii) without the inverse covariance between the two derivatives} (see Eq. \ref{eq:Fisher_Gaussian}). Both of these simplifications were needed in order to stabilize the training of the IMNN, and they are correct as long as the covariance is kept at unity.\footnote{The original work by \cite{Charnock2018} had slightly different loss function, by having the inverse of the covariance in the Fisher term and the additional Frobenius norm between $\bm{\Sigma}^{-1}$ and $\bm{\mathbb{1}}$. We find that these choices destabilized the training and are not crucial for the convergence (see also \citealt{Makinen2021}).}

For the IMNN training, we created a database around the fiducial $\bm{\theta}_{\text{fid}}$ (see Figs. \ref{fig:pipeline} and \ref{fig:scatter_plot}) for $N = M = 1024$ (Eqs. \ref{eq:IMNN_mu}, \ref{eq:IMNN_sigma}, \ref{eq:IMNN_dmu}), with $11 264$ realizations in total. The training and validation sets were formed by splitting the database in half, following the original work of \citet{Charnock2018}. We trained the network using stochastic gradient descent with a batch size $b = 128$, regularization strengths $\lambda = 10$, $\lambda_W = 10^{-9}$ and the \texttt{Adam} optimizer. To increase the effective database size and improve the training, the telescope noise was added to the simulations \enquote{on the fly.} The final model has been trained on $8$ V-100 GPUs in parallel, for $\approx 120\mathrm{k}$ epochs and $\approx 20\mathrm{k}$ GPUh. The IMNN code developed for the purposes of this work we make publicly available under \texttt{21cmIMNN}\footnote{\url{https://github.com/dprelogo/21cmIMNN}} package.

\subsection{2DPS + IMNN}
Finally, we also computed the Fisher information of the concatenation of the individual 2DPS and IMNN summaries.  As we shall see in the following section, these are the two best performing individual summaries. We note that one can cleanly combine summaries when performing simulation-based inference (SBI), as the likelihood does not need to be specified explicitly.  The main restriction is the size of the required database and the stability of the SBI training (e.g., \citealt{Cranmer2020, Lueckmann2021}). When computing the Fisher information in this work, we were restricted to these two summaries due to the limits imposed by our database size. In particular, the minimal number of samples needed to estimate the covariance matrix is proportional to the dimensionality of the total summary (e.g., \citealt{Hartlap2007}). Furthermore, the dimensions of the final summary should be uncorrelated, as otherwise the covariance is singular and non-invertible. This can result in incorrect Fisher matrix estimation (see however \citealt{Tegmark1997} for possible alternative solutions).

\section{Database}
\label{sec:database}
\begin{figure}
    \centering
    \includegraphics[width=\linewidth]{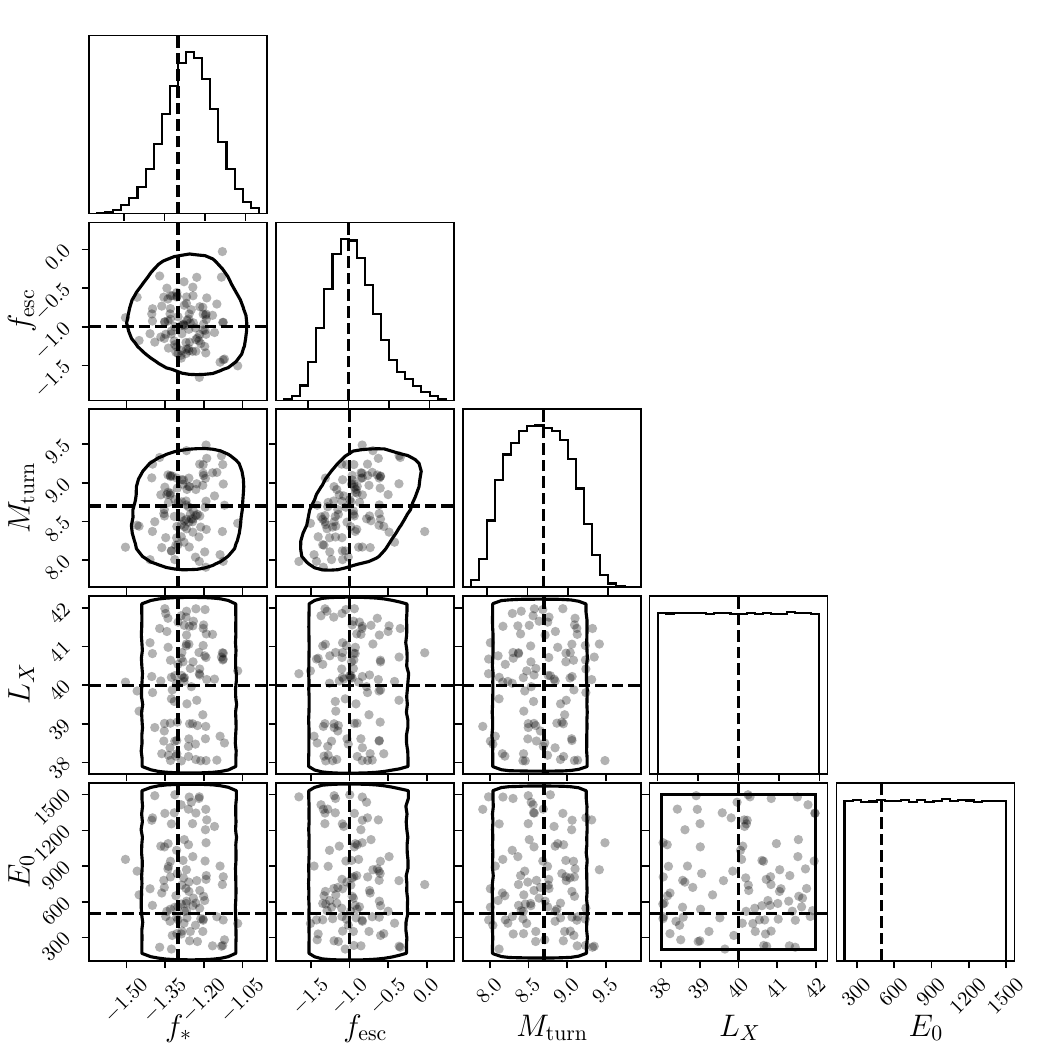}
    \caption{Prior volume of the astrophysical parameter space. Dashed lines demarcate the fiducial parameter values used to train the IMNN, 2D contours enclose 95\% of the prior volume, and the points denote the 152 prior samples at which we computed the Fisher information for our summaries.  Note that all units and $\log_{10}$ factors have been removed for clarity; see the main text and Fig. \ref{fig:pipeline} for reference on the units.}
    \label{fig:scatter_plot}
\end{figure}

Since we did not have a strongly motivated fiducial choice for our galaxy parameters, we wanted to compare the Fisher information of 21 cm summaries across our prior volume.  To do so, we constructed a \enquote{Fisher database,} sampling  152 parameter combinations from our prior.  We used a flat prior for the X-ray emission parameters, $L_X$ and $E_0$, over the ranges shown  in Fig. \ref{fig:scatter_plot} (for more motivation of these choices see \citealt{Fragos2013, Lehmer2016, Das2017}).  The remaining parameters characterizing star formation and ionizing emission of galaxies are constrained by observed UV luminosity functions (e.g.,  \citealt{Bouwens2015b, Bouwens2017, Oesch2018}) as well as estimates of the reionization history (e.g., \mbox{\citealt{McGreer2015}}; \mbox{\citealt{Planck2016XLVII}}).  As our model is nested in the 6D \cite{Park2019} model, we used the posterior from that work, calculating the corresponding conditional distributions over $f_{*, 10}$, $f_{\text{esc}, 10}$ and $M_{\text{turn}}$, fixing the remaining parameters to their fiducial values, as discussed in the previous section.  To compute the conditional distributions, we used the \texttt{conditional\_kde}\footnote{\url{https://github.com/dprelogo/conditional_kde}} code.

Our prior and samples are shown in Fig. \ref{fig:scatter_plot}.  The diagonal panels illustrate the 1D prior probability distribution functions (PDFs), while the contours in the lower left enclose 95\% of the 2D distributions.  The 152 samples are shown as points, while the dashed vertical and horizontal lines demarcate the fiducial choice used to train our IMNN.

We simulated $N = 130$ realizations for each of the 152 parameter samples, $\bm{\theta^*}$, in order to estimate the mean and the covariance of each summary (Eqs. \ref{eq:mu_mean}, \ref{eq:sigma_mean}), plus an additional $2 \times 5 \times M = 150$ realizations at $\bm{\theta^*} \pm \delta\bm{\theta}_m / 2$ to estimate the derivative of the mean (Eq. \ref{eq:mu_der}) along each of the 5 parameter basis vectors.
We note that no new simulations are needed  to estimate the derivative of the covariance matrix, since we used the same  $L = M = 15$ realizations that were used to compute the first Fisher term mentioned above (c.f. Eq. \ref{eq:sigma_der} and associated discussion). In Appendix \ref{app:convergence} we show convergence tests of these choices for estimating the Fisher information of our summaries. {  We caution that the WST and RNN summaries require a larger dataset to fully converge.}

\section{Results} \label{sec:results}
We now present the main results of this work.  We  first discuss the training of the IMNN in Section \ref{sec:IMNN}, then compare the performance of all of our 21 cm summaries in Section \ref{sec:compare}, adding a final discussion on the importance of including the second Fisher term (c.f. Eq. \ref{eq:Fisher_Gaussian}) in Section \ref{sec:second_term}.  Throughout, our figure of merit will be the PDF of the Fisher information ($\log_{10} (\det \bm{F})$), over the prior volume (see Sect. 4).  We quote both the median and the variance of this distribution.  A high value of the median means that the typical parameter value is well constrained by the summary, while a narrow variance means that the constraining power of the summary is relatively constant over the prior volume.  We note that the Fisher information is a measure of the {average} constraining power over our five parameters; we do not discuss the relative constraints on individual parameters in this work.

\subsection{IMNN training}
\label{sec:IMNN}
\begin{figure}
    \centering
    \includegraphics[width=\linewidth]{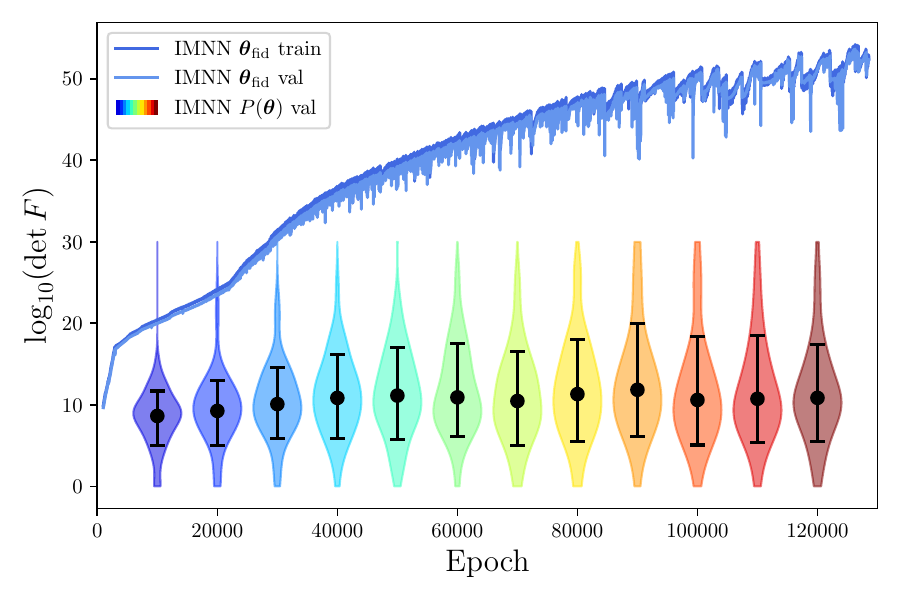}
    \caption{ Fisher information of the IMNN during training.  Dark (light) blue curves show the Fisher information for the training (validation) set of the IMNN trained at $\bm{\theta}_{\rm fid}$.  The violin plots show the distribution of the Fisher information over the full prior volume excluding $\bm{\theta}_{\rm fid}$, $P(\theta)$, at every interval of the 10k training epochs.  Black dots and error bars inside each violin denote the median and 68\% CLs of the distributions. The divergence between the Fisher information computed at $\bm{\theta}_{\rm fid}$ and its PDF over the remaining prior volume clearly indicates that the IMNN is over-specializing to the specific parameter combination, $\bm{\theta}_{\rm fid}$, used in the training.}
    \label{fig:IMNN_training_violins}
\end{figure}

In Fig. \ref{fig:IMNN_training_violins} we show the training results for the IMNN. Blue lines show the training and validation Fisher information (see the first term of the IMNN loss, Eq. \ref{eq:IMNN_loss}) at the fiducial point $\bm{\theta}_{\text{fid}}$.  We recall that the training and validation sets each consist of 5632 light cone realizations, evaluated at $\bm{\theta}_{\text{fid}}$ with different cosmic initial conditions and telescope noise seeds. The Fisher information of the training and validation sets are largely overlapping meaning the network does not overfit to the training set.  We see that by the end of the training, the IMNN has learned to improve the Fisher information by 40(!) orders of magnitude, corresponding to parameter constraints that are improved by an average factor of $\sim10^8$ in the variance!

Is the compression learned by the IMNN comparably informative at other parameter combinations?  We can answer this question by comparing the solid lines of Fig. \ref{fig:IMNN_training_violins} to the violin distributions.  The latter show the distribution of the Fisher information over the full prior volume, $P(\theta)$, at every interval of 10k training epochs, excluding $\bm{\theta}_{\rm fid}$.
We see that although the IMNN is not overfitting in the classical sense (the Fisher information of the validation set keeps increasing), it quickly \enquote{over-specializes} to the specific parameter value used for training, $\bm{\theta}_{\rm fid}$.   Thus, the compression learned by the IMNN does not generalize well across parameter space.\footnote{One solution could be to train the IMNN around several different points. However, this is extremely expensive in our application, both in terms of database size and GPU time. Furthermore, it is not clear if such a procedure would result in a meaningful summary, as the concept of one point training is inherently built into the IMNN. Alternatively, it might be possible to construct a generalized IMNN, maximizing the Fisher information throughout parameter space; however, it is not obvious how this could be done.}
The PDF of the Fisher information over the prior volume does not change significantly after epoch 40k, with the median beginning to decrease beyond epoch 90k.  We took epoch 90k as the final state of the IMNN in the remainder of this work.

\subsection{Comparison of all summaries}
\label{sec:compare}
\begin{figure}
    \centering
    \includegraphics[width=\linewidth]{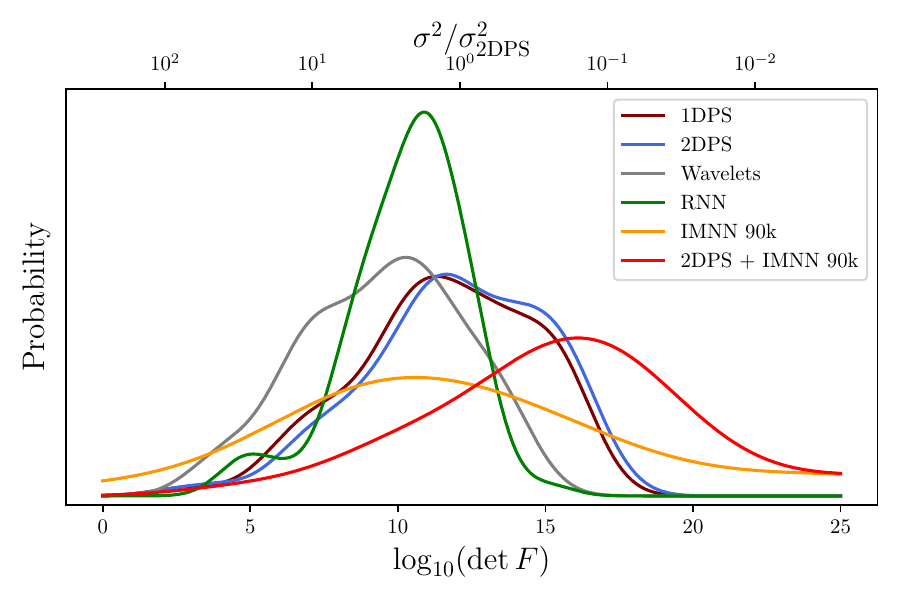}
    \caption{Distribution of the Fisher information ($\det \bm{F}$) across the prior volume, for all summaries considered in this work. For the IMNN summary, we used the weights that give the highest median value (corresponding to 90k epochs; see Fig. \ref{fig:IMNN_training_violins}).  On the top axis we denote the corresponding relative improvement in the {single parameter variance} (Eq. \ref{eq:single_parameter_variance}), normalized to its median value of the 2DPS summary. 
    The 2DPS has the largest median value of all of the individual summaries, while the combination of the 2DPS + IMNN dramatically outperforms every individual summary.  See Table \ref{tab:results} for quantitative values.}
    \label{fig:all_in_one}
\end{figure}
\begin{table}
\caption{\label{tab:results} Distributions of the Fisher information.}
\centering
\begin{tabular}{lcc}
\vspace{0.1cm} summary  & $\text{median}^{+34\%}_{-34\%}$ &  $\sigma^2 / \sigma^2_{\text{2DPS}}$ \\ \hline \\
\vspace{0.25cm} Wavelets           & $\phantom{0}9.6_{-3.1}^{+2.9}$  & 3.16\\
\vspace{0.25cm} {  RNN}                & {  $10.5_{-2.2}^{+1.8}$}  & {  2.09}\\
\vspace{0.25cm} IMNN 90k           & $11.8_{-5.7}^{+8.2}$ & 1.15\\
\vspace{0.25cm} 1DPS               & $11.8_{-3.4}^{+3.1}$ & 1.15\\
\vspace{0.25cm} 2DPS               & $12.1_{-3.3}^{+3.1}$ & 1.00\\
\vspace{0.25cm} IMNN 90k + 2DPS    & $15.9_{-4.9}^{+5.2}$ & 0.17
\end{tabular}
\tablefoot{Companion table to Fig. \ref{fig:all_in_one}, summarizing the Fisher information probability distribution across the prior volume for each summary. In particular, the first column represents values of $\log_{10} (\det \bm{F})$ for the median and $68\%$ of probability volume around it. The second column shows the {single parameter variance} (Eq. \ref{eq:single_parameter_variance}), which serves as a guide to how much would the recovered variance of a parameter increase or decrease if one uses that particular summary instead of the 2DPS.}
\end{table}

In Fig. \ref{fig:all_in_one} we show the main result of this work: the distribution of Fisher information across the prior volume, for all of the summary statistics we consider. On the top axis we denote the corresponding relative improvement in constraints on the {single parameter variance} $\sigma^2$. We define it as an effective variance in each dimension with no correlation, giving the same Fisher information. For a 5D parameter space we have
\begin{equation}
    \sigma^2 = (\det \bm{F})^{-1 / 5} \, . \label{eq:single_parameter_variance}
\end{equation}
Here, one can ignore the difference in the units (i.e., scaling) between parameter dimensions (see Appendix \ref{app:Fisher_transformation} for more details). Specifically, the top axis of Fig. \ref{fig:all_in_one} denotes the single parameter variance normalized to its median value of the 2DPS summary. To give an example, a factor of $10$ times larger Fisher information for a 5-dim parameter space would result in a $37\%$ smaller single parameter variance. The median values and 68\% confidence levels (CLs) of the Fisher information and single parameter variance are also listed in Table \ref{tab:results}.

Among the individual summaries, the 2DPS is a clear winner.  In comparison, the 1DPS has roughly a factor of 2 smaller Fisher information across the prior volume, which translates to a $\sim15\%$ larger single parameter variance (Eq. \ref{eq:single_parameter_variance}). This {  difference between the 1DPS and 2DPS}  shows that indeed the anisotropy of the cosmic 21 cm signal helps in parameter recovery, though not very significantly. 
The distribution of the 2DPS Fisher information is also relatively narrow, indicating its constraining power does not vary enormously across parameter space, compared to most other summaries.

The median Fisher information of the IMNN is comparable with that of the 1DPS.  However, the IMNN results in the widest distribution of all of the summaries.  This means that the compression that the IMNN learned at $\bm{\theta}_{\rm fid}$ can be much more informative but also much less informative, depending on the parameter choice.  Therefore, if one has a good idea of where the maximum likelihood value will be (e.g., from complementary observations; \citealt{Park2019, HERA2022b, HERA2023, Breitman2024}), it could be beneficial to train the IMNN at that {best guess} value and use the resulting summary in inference.  An optimal strategy could be to perform inference using the 2DPS in order to obtain the maximum likelihood estimate (MLE), train the IMNN at this MLE parameter combination, then perform inference again using the resulting IMNN summary.

The RNN performs worse than the two PS summaries, resulting in a factor of 2 larger single parameter variance compared to the 2DPS. 
However, its distribution is the most homogeneous across the prior volume (i.e., the narrowest). This could be interpreted as sacrificing the overall information content for robustness throughout the parameter space. 
The RNN was trained as a regressor over a large parameter space, and so it is understandable that it is much more consistent with respect to the IMNN, which was trained on a single parameter combination.  The overall low median information content of the RNN compression could be due to the fact that it was trained on a different parametrization of the cosmological signal than used in this work.  We expect that retraining the RNN as a regressor on the same parametrization would improve the median information content.  Indeed, \cite{Zhao2022a} find that a regressor CNN slightly outperforms the 1DPS, resulting in tighter parameter constraints when using two different parameter combinations for the mock observation.   However, it is likely that this increase in the information content would come at the cost of robustness (i.e., a broader PDF), as the RNN over-specializes to features inherent in that model's parametrization and its training set. {  Additionally, we caution the Fisher estimate for the RNN summary needs a larger dataset to fully converge (c.f. Appendix \ref{app:convergence}).}

The wavelet summary results in a median Fisher information that is a factor of $\sim$300 smaller than that of the 2DPS at their median values (factor of $\sim$3 increase in the single parameter variance (Eq. \ref{eq:single_parameter_variance}).  Intuitively, one would expect more information content in wavelets compared with the PS, as they effectively include higher-order correlations. However, our WST followed the definition in \cite{Greig2022, Greig2023}, in which the wavelets are only computed at one slice for each redshift chunk.  Thus, the physics-rich, line-of-sight modes inside each redshift bin are lost for this choice of WST.   The weaker performance of the WST { (as defined in this work)} compared to the power spectra implies that there is more information in the line-of-sight modes than there is in higher-order correlations of the transverse modes.

We note that recent works also using 2D wavelets achieved better results than we do here \citep{Greig2022, Hothi2023}.  However, their analysis differs from ours. In particular, they use more redshift bins and they do not average the coefficients over the sky-plane (rolling the filter as described above).  We confirm that without the additional smoothing that we performed in this work, the Gaussian approximation for the likelihood that is intrinsic to the Fisher estimate (c.f. Eq. \ref{eq:Fisher_Gaussian}) is notably worse, for both the wavelets and the RNN.  Therefore, the higher wavelet information found by \citet{Greig2022, Hothi2023} could be partially due to a less appropriate application of the Fisher estimate.
Alternatively, the information content could be improved by using 3D wavelets instead of 2D ones (see, e.g., \citealt{Zhao2023b}). {  As for the RNN, we caution that the WST needs a larger dataset for the Fisher to fully converge (c.f. Appendix \ref{app:convergence}).}

Finally, our concatenation of the 2DPS and IMNN significantly outperforms all of the individual summaries (c.f. the red curve in Fig. \ref{fig:all_in_one}).  As we saw above, these two summary statistics are highly complimentary: the 2DPS is robust (i.e., a narrow information PDF), while the IMNN summary can be much more constraining for some parameters but much less for others (i.e., it has a very broad information PDF).
The distribution of Fisher information of 2DPS + IMNN retains the narrowness of the 2DPS, but is shifted to the right, including the high-value tail from the IMNN. 
\footnote{We note that in the case in which some value of the IMNN summary would be completely degenerate with some 2DPS bin, the resulting covariance matrix would be singular and therefore non-invertible. This would make estimating the Fisher matrix unreliable. However, as the IMNN compression and 2DPS are very different summaries, this is unlikely to be a problem.} The combined summary is a factor of $\sim$6.5--9.5 better in the single parameter variance (Eq. \ref{eq:single_parameter_variance}), averaged over the prior samples. Combining multiple 21 cm statistics has already been proven to tighten parameter constraints (e.g., \citealt{Gazagnes2021, Watkinson2022}).  However, previous work using traditional inference was limited by the need to derive a tractable likelihood, which becomes impossible for nontrivial summaries.  Fortunately, recent advances in applying SBI to 21 cm (e.g.,\mbox{\citealt{Zhao2022a, Zhao2022b};} \citealt{Prelogovic2023, Saxena2023}) will allow us to cleanly combine multiple summary statistics without having to explicitly define the likelihood.   Unfortunately, the neural density estimation that is intrinsic to SBI can be unstable and/or require large training sets, if the summary statistics become very high dimensional.  This work illustrates how combining only two complimentary statistics can result in much tighter, more robust parameter constraints than would be available using either individually.

\subsection{One-term versus full Fisher matrix}
\label{sec:second_term}

\begin{figure}
    \centering
    \includegraphics[width=\linewidth]{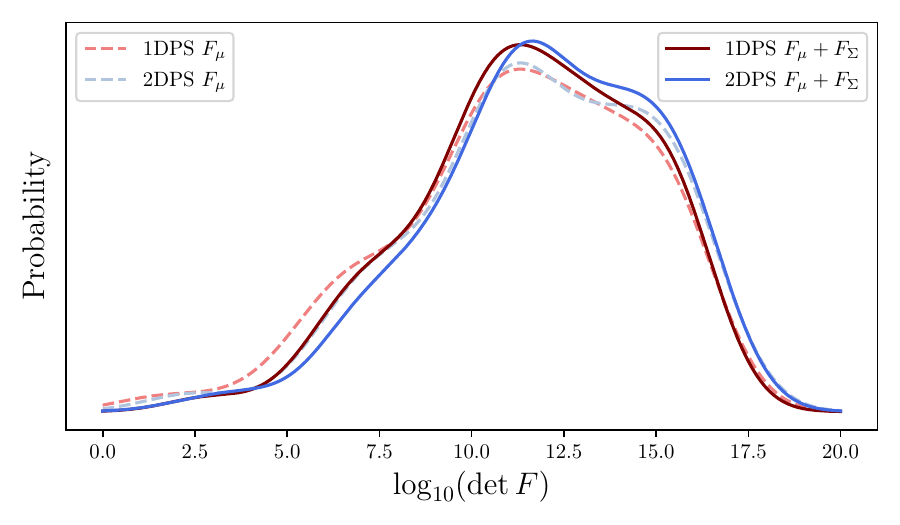}
    \caption{Distributions of Fisher information for the 1DPS (red) and 2DPS (blue).  Solid lines account for both terms in the Fisher matrix (c.f. Eq. \ref{eq:Fisher_Gaussian}), while the dashed lines correspond to the common simplification of a constant covariance (i.e., including only the first term in Eq. \ref{eq:Fisher_Gaussian}).  The assumption of a constant covariance underpredicts the Fisher information by a factor of $\sim 2$. For a 5-parameter model, this translates to a $\sim15\%$ larger single parameter variance (Eq. \ref{eq:single_parameter_variance}).}
    \label{fig:1D_vs_2D}
\end{figure}
\begin{figure}
    \centering
    \includegraphics[width=\linewidth]{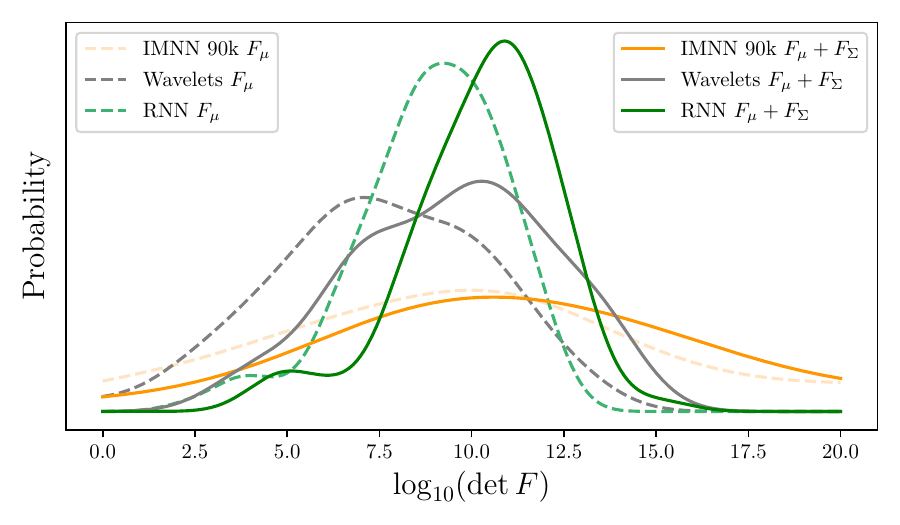}
    \caption{Analogous to Fig. \ref{fig:1D_vs_2D}, but for the other 21 cm summary statistics.}
    \label{fig:one_term_vs_full_fisher}
\end{figure}

The full Fisher matrix of a Gaussian likelihood can be separated in two terms $\bm{F} = \bm{F}_{\bm{\mu}} + \bm{F}_{\bm{\Sigma}}$ (see Eq. \ref{eq:Fisher_Gaussian}), where the first term includes the gradient of the mean and the second term the gradient of the covariance matrix (e.g., Appendix \ref{app:gaussian_fisher}; \citealt{Tegmark1997}). {  All of our results from the previous sections were calculated using the full expression, including both terms.}  However, it is common in the literature to ignore one of the terms, by approximating the mean or the covariance matrix as a constant (e.g., \citealt{Carron2013}).  In most of cosmological Fisher forecasts using the PS (e.g., \citealt{Sailer2021, Abazajian2022, Euclid2022, Mason2023, Bykov2023}), the second term, $\bm{F}_{\bm{\Sigma}}$, is neglected.\footnote{For examples of ignoring the first term, see, e.g., \cite{Abramo2012} and \cite{dAssignies2023}.}
In some cases, the covariance matrix can indeed be independent or only weakly dependent on the parameters, justifying ignoring the second Fisher term.
However, the importance of the $\bm{F}_{\bm{\Sigma}}$ term for 21 cm Fisher forecasts has not yet been investigated (e.g., \citealt{Greig2022, Balu2023, Mason2023, Hothi2023}).  Indeed in \citet{Prelogovic2023} we showed that assuming a parameter-independent covariance matrix of the 21 cm 1DPS likelihood can bias the posterior estimation.  In this section, we quantify the importance of the $\bm{F}_{\bm{\Sigma}}$ term in determining the Fisher information\footnote{Assuming a constant covariance matrix can result in other biases at the level of the full Fisher matrix.  For example, it could impact parameter degeneracies and/or improve constraints of some parameters at the cost of others. Here we only focus on the effect this simplification has on the determinant of the Fisher matrix as this quantity measures the average parameter constraining power of the summary statistics.}, for all of our 21 cm summaries. It is important to note that adding this term will always increase the information content, as $\det (\bm{F}_{\bm{\mu}} + \bm{F}_{\bm{\Sigma}}) \ge \det \bm{F}_{\bm{\mu}}$. More details are given at the end of Appendix \ref{app:det_proof}.

In Fig. \ref{fig:1D_vs_2D} we plot the Fisher information PDF of the commonly used 1DPS and 2DPS.  The solid curves are the same as in the previous figure, corresponding to the full expression for the Fisher information.  The dashed curves only include the first term, assuming a constant covariance matrix. For both 1D and 2D power spectra, neglecting the second term underestimates the Fisher information by a factor of $\sim2.3$ on average over the prior samples.  
For our 5-parameter model, this translates into an underestimate of the single parameter variance (Eq. \ref{eq:single_parameter_variance}) by 15\%.

In Fig. \ref{fig:one_term_vs_full_fisher} we compare the Fisher information PDFs with and without the second Fisher term for the remaining individual summaries.  We see that for the other 21 cm summaries, the second Fisher term is even more important than for the power spectra.  For example, assuming a constant covariance underestimates the mean Fisher information of the RNN for a factor of {  $\sim$2000} and a factor of $\sim$11100 for the wavelets, averaged over the prior samples. We note that although the IMNN is trained to normalize the covariance to unity, thus ensuring a constant covariance matrix by construction, this is only done at $\bm{\theta}_{\text{fid}}$.  For the other points in the parameter space, there is no guarantee that this condition is met.

\section{Conclusions} \label{sec:conclusions}
The cosmic 21 cm signal will revolutionize our understanding of the CD and EoR.  "Big Data" analysis techniques will be required to cope with radio telescopes such as the SKA.   In this work we compare several approaches to compressing 21 cm light cone data, on the basis of their Fisher information.  Specifically, we compare the determinant of the Fisher information for the following summary statistics:  (i) the 1DPS, (ii) the 2DPS, (iii) the 2D WST, (iv) a RNN trained as a regressor; (v) an IMNN; and (vi) the combination of 2DPS and IMNN.  Importantly we compare their Fisher information across the parameter space since there is currently no standard model for the astrophysics of the first galaxies, which determines the cosmic 21 cm signal.

The 2DPS is the individual summary with the highest median Fisher information (Fig. \ref{fig:all_in_one}, Table \ref{tab:results}). 
The distribution of its Fisher information is also relatively narrow across the parameter space.  This means that compared to most other summaries, the 2DPS can recover astrophysical parameters with relatively similar precision regardless of what the "true" values of our Universe are.
We also show that the 2DPS slightly outperforms the 1DPS, confirming that the anisotropy of the 21 cm signal helps constrain parameters.
The information content of both the RNN and wavelets is generally lower than either the 1DPS or the 2DPS.  However, we caution that both of these summaries could be defined in several different ways, potentially improving their performances (e.g., \citealt{Greig2022, Greig2023, Hothi2023, Zhao2023b}).  Moreover,  our Fisher estimates for the RNN and wavelet summaries require a larger dataset to fully converge (see Appendix \ref{app:convergence}).

In this work we also introduce IMNNs to the field of 21 cm cosmology.  Although capable of achieving the highest Fisher information for some parameter choices, the IMNN does not generalize well, resulting in a broad distribution across the prior volume. 
We find an enormous difference between its Fisher information computed at the parameter set used for training, $\bm{\theta}_{\text{fid}}$, and its information computed at the remaining prior samples (Fig. \ref{fig:IMNN_training_violins}).  This means that the IMNN overfits to the information content of the fiducial. The regularization inherent to the IMNN loss (see Eq. \ref{eq:IMNN_loss} and the following discussion) makes it difficult to avoid this overfitting by training over many different choices of $\bm{\theta}_{\rm fid}$.  A simpler solution could be to first perform inference with a less optimal summary, in order to determine a maximum likelihood point, and then train the IMNN at that point.

Combining the IMNN with the 2DPS results in a summary whose median Fisher information is almost $\sim$4 orders of magnitude larger than the 2DPS alone. This results in a factor of $\sim$6.5--9.5 tighter single parameter variance averaged over the prior samples. This means the IMNN extracts complementary information with respect to the 2DPS. Performing SBI using such a summary could yield extremely tight and robust parameter constraints.  Moreover, combining only these two complementary summaries still preserves a high level of compression, with our light cone of $\sim 1.6 \cdot 10^6$ cells reduced to just 77 numbers.

Finally, we stress the importance of not assuming a constant covariance when performing Fisher forecasts using 21 cm observables.  This common assumption  underestimates the Fisher information of the PS on average by a factor of $\sim$2.3, corresponding to a 15\% increase in the single parameter variance (Eq. \ref{eq:single_parameter_variance}).  For other summaries, the assumption underestimates the Fisher information by orders of magnitude. 

\begin{acknowledgements}
D.P. thanks T. Charnock for many discussions and in depth explanations on how the IMNN function, L. Makinen on help with stabilizing IMNN training, and C. Sui for the discussion on Mutual Information. We thank B. Greig and S. Murray for many useful discussions and for giving the comments on the manuscript. A.M. acknowledges support from the Ministry of Universities and Research (MUR) through the PRIN project "Optimal inference from radio images of the epoch of reionization" as well as the PNRR project "Centro Nazionale di Ricerca in High Performance Computing, Big Data e Quantum Computing".   We  gratefully acknowledge computational resources of the HPC center at SNS. We acknowledge the CINECA award under the ISCRA initiative, for the availability of high performance computing resources and support (grant IMC21cm - HP10CWIEF7). This work used the Extreme Science and Engineering Discovery Environment (XSEDE), which is supported by National Science Foundation grant number ACI-1548562. Specifically, it used the Bridges-2 system, which is supported by NSF award number ACI-1928147, at the Pittsburgh Supercomputing Center (PSC).
\end{acknowledgements}

\bibliographystyle{aa} 
\bibliography{references} 

\begin{appendix}
\section{Proof of the determinant inequality} \label{app:det_proof}
We considered two matrices, $\bm{C}$ and $\bm{F}^{-1}$, of shape $m\times m$ for which the following inequality holds:
\begin{equation}
    \bm{C} \ge \bm{F}^{-1} \, .
\end{equation}
This means that their difference is a  positive, semi-definite matrix:
\begin{equation}
    \bm{x}^T (\bm{C} - \bm{F}^{-1}) \bm{x} \ge 0 \quad \forall \bm{x} \in \mathbb{R}^{m} \, .
\end{equation}
From this, it follows that
\begin{align}
    \bm{x}^T \bm{C} \bm{x} &\ge \bm{x}^T \bm{F}^{-1} \bm{x} \\
    \exp\left(-\frac{1}{2}\bm{x}^T \bm{C} \bm{x}\right) &\le \exp\left(-\frac{1}{2}\bm{x}^T \bm{F}^{-1} \bm{x}\right) \\
    \int\mathrm{d}\bm{x}\, \exp\left(-\frac{1}{2}\bm{x}^T \bm{C} \bm{x}\right) &\le \int\mathrm{d}\bm{x}\, \exp\left(-\frac{1}{2}\bm{x}^T \bm{F}^{-1} \bm{x}\right) \\
    \frac{1}{\sqrt{\det (\bm{C} / 2\pi)}} &\le \frac{1}{\sqrt{\det (\bm{F}^{-1} / 2\pi)}} \\
    \det \bm{C} &\ge \det \bm{F}^{-1}.
\end{align}
Here the trick was to push the inequality toward the multivariate Gaussian integral. After integrating, what is left is only the inverse of a normalization constant. We note that if one wants to interpret $\bm{C}$ as a covariance matrix and $\bm{F}$ as a Fisher matrix, then it would make more sense to integrate over $\bm{x}^T \bm{C}^{-1} \bm{x}$ and  $\bm{x}^T \bm{F} \bm{x}$. However, this is irrelevant for the proof, as it would only result in the change of the inequality direction in each row, ending with $\det \bm{C}^{-1} \le \det \bm{F}$. We usually assume the matrices are nonsingular and thus invertible, which is needed here. 

We note that the identical procedure can be used to prove that the one-term Fisher is a lower bound of the two-term Fisher. For $\bm{F} = \bm{F}_{\bm{\mu}} + \bm{F}_{\bm{\Sigma}}$ one can write $\bm{x}^T (\bm{F}_{\bm{\mu}} + \bm{F}_{\bm{\Sigma}}) \bm{x} \ge \bm{x}^T \bm{F}_{\bm{\mu}} \bm{x}$, which by following the same steps translates into $\det (\bm{F}_{\bm{\mu}} + \bm{F}_{\bm{\Sigma}}) \ge \det \bm{F}_{\bm{\mu}}$. 

\section{Derivation of a Gaussian Fisher information matrix} \label{app:gaussian_fisher}
To derive Eq. \ref{eq:Fisher_Gaussian} (i.e., the Fisher information of a multivariate Gaussian distribution), we started by writing the Fisher information matrix equation (Eq. \ref{eq:Fisher_def}) in the equivalent form:
\begin{equation}
    \bm{F}(\bm{\theta}^*)_{mn} = - \mathbb{E}\left[\left.\partial^2_{mn} \ln P(\bm{d} | \bm{\theta})\right|\bm{\theta^*}\right] \, , \label{eq:app:equivalent_fisher}
\end{equation}
where we shorten the notation of derivatives as $\partial/\partial\bm{\theta}_m \equiv \partial_m$ and $\partial^2/\partial\bm{\theta}_m\partial\bm{\theta}_n \equiv \partial^2_{mn}$. This form  comes from the fact that 
\begin{equation}
    - \partial^2_{mn} \ln P(\bm{d}|\bm{\theta}) = \partial_m \ln P(\bm{d}|\bm{\theta}) \cdot \partial_n \ln P(\bm{d}|\bm{\theta}) - \frac{1}{P(\bm{d}|\bm{\theta})} \partial^2_{mn} P(\bm{d}|\bm{\theta}) \, .
\end{equation}
Taking the expectation value on both sides, together with
\begin{equation}
    \mathbb{E}\left[\left.\frac{1}{P(\bm{d}|\bm{\theta})} \partial^2_{mn} P(\bm{d}|\bm{\theta}) \right|\bm{\theta}^*\right] = \partial^2_{mn} \int \mathrm{d}\bm{d} \, P(\bm{d}|\bm{\theta}^*) = 0 \, , 
\end{equation}
proves the equality between Eqs. \ref{eq:Fisher_def} and \ref{eq:app:equivalent_fisher}.

Before diving into evaluating \ref{eq:app:equivalent_fisher} for a Gaussian likelihood $\mathcal{N}(\bm{d} | \bm{\mu}(\bm{\theta}), \bm{\Sigma}(\bm{\theta}))$, we firstly list several useful relations:
\begin{align}
    \partial_m \bm{\Sigma}^{-1} &= - \bm{\Sigma}^{-1} \, \partial_m \bm{\Sigma} \, \bm{\Sigma}^{-1} \, , \\
    \partial_m |\bm{\Sigma}| &= |\bm{\Sigma}| \,  \operatorname{tr} \left(\bm{\Sigma}^{-1} \partial_m \bm{\Sigma}\right) \, , \\
    \mathbb{E}\left[\bm{A} (\bm{d} - \bm{\mu})\right] &= \bm{0} \, ,  \label{eq:app:E1}\\
    \mathbb{E}\left[(\bm{d} - \bm{\mu})^T \bm{A} (\bm{d} - \bm{\mu})\right] &= \operatorname{tr}(\bm{A\Sigma}) \, . \label{eq:app:E2}
\end{align}
The first one follows from $\bm{\Sigma}\bm{\Sigma}^{-1} = \mathbb{1}$ and calculating its derivative, the second one is Jacobi's relation for the derivative of the determinant. The last two are expectation relations for Gaussian distributions for an arbitrary matrix $\bm{A}$ (where $\dim \bm{A} = \dim \bm{\Sigma}$). The first of these follows from the antisymmetry of the integral, and the second is left as an exercise for the conscientious reader. 

With this at hand, we can attack the problem of calculating the Fisher matrix for a Gaussian likelihood. By taking the logarithm
\begin{equation}
    \ln P(\bm{d}|\bm{\theta}) = \text{const.} - \frac{1}{2} \ln |\bm{\Sigma}| - \frac{1}{2}(\bm{d} - \bm{\mu})^T \bm{\Sigma}^{-1} (\bm{d} - \bm{\mu})
\end{equation}
and computing its second derivative, we get
\begin{align}
    \partial^2_{mn} \ln P(\bm{d}|\bm{\theta}) &= - \partial_m\bm{\mu}^T \, \bm{\Sigma}^{-1} \partial_n\bm{\mu} \\
    &\phantom{=} + \frac{1}{2} \operatorname{tr} \left(\bm{\Sigma}^{-1} \, \partial_m \bm{\Sigma} \, \bm{\Sigma}^{-1} \, \partial_n \bm{\Sigma} \right) \\
    &\phantom{=} - (\bm{d} - \bm{\mu})^T \, \bm{\Sigma}^{-1} \, \partial_m \bm{\Sigma} \, \bm{\Sigma}^{-1} \, \partial_n \bm{\Sigma} \, \bm{\Sigma}^{-1} (\bm{d} - \bm{\mu}) \\
     &\phantom{=} - \frac{1}{2}\operatorname{tr} \left(\bm{\Sigma}^{-1} \, \partial^2_{mn} \bm{\Sigma} \right)\\
     &\phantom{=} + \frac{1}{2}(\bm{d} - \bm{\mu})^T \, \bm{\Sigma}^{-1} \, \partial^2_{mn} \bm{\Sigma} \, \bm{\Sigma}^{-1} (\bm{d} - \bm{\mu}) \\
     &\phantom{=} + \partial^2_{mn}\bm{\mu}^T \bm{\Sigma}^{-1} (\bm{d} - \bm{\mu}) \\
     &\phantom{=} - \left(\partial_m\bm{\mu}^T \bm{\Sigma}^{-1} \partial_n\bm{\Sigma} + \partial_n\bm{\mu}^T \bm{\Sigma}^{-1} \partial_m\bm{\Sigma}\right) \bm{\Sigma}^{-1} (\bm{d} - \bm{\mu}).
\end{align}
Taking the negative expectation value of the final equation can  now be done by using relations \ref{eq:app:E1} and \ref{eq:app:E2}. One can easily see that only the first three terms contribute, with others canceling each other out or being equal to zero. With this, we finally have\begin{equation}
    \bm{F}_{mn} = \partial_m\bm{\mu}^T \, \bm{\Sigma}^{-1} \partial_n\bm{\mu} + \frac{1}{2} \operatorname{tr} \left(\bm{\Sigma}^{-1} \, \partial_m \bm{\Sigma} \, \bm{\Sigma}^{-1} \, \partial_n \bm{\Sigma} \right) \, .
\end{equation}

\section{Fisher information under linear transformations} \label{app:Fisher_transformation}
Imagine we calculate the Fisher information matrix in parameter space $\bm{\theta}$ around $\bm{\theta^*}$:
\begin{equation}
        \bm{F}_{\bm{\theta}}(\bm{\theta^*})_{mn} = \mathbb{E}\left[\left.\frac{\partial}{\partial\bm{\theta}_m} \ln P(\bm{d} | \bm{\theta}) \cdot \frac{\partial}{\partial\bm{\theta}_n} \ln P(\bm{d} | \bm{\theta})\right|{\bm{\theta^*}} \right] \, .
\end{equation}
If we now take a linear transformation of the parameter space $\bm{\eta} = \bm{a} \odot \bm{\theta} + \bm{b}$, where $\odot$ is element-wise multiplication, we get
\begin{equation}
    \bm{F}_{\bm{\eta}}(\bm{\eta}^* = \bm{a} \odot \bm{\theta}^* + \bm{b}) = \bm{A} \, \bm{F}_{\bm{\theta}}(\bm{\theta^*}) \, \bm{A} \, ,
\end{equation}
where $\bm{A} = \operatorname{diag}(\bm{a})$ is a diagonal matrix with elements on the diagonal equal to the vector $\bm{a}$. That further means:
\begin{align}
    \det\bm{F}_{\bm{\eta}} &= \det{\bm{A}  \bm{F}_{\bm{\theta}} \bm{A}} = (\det\bm{A})^2 \, \det\bm{F}_{\bm{\theta}} \, , \\
    \ln \det\bm{F}_{\bm{\eta}} &= C + \ln\det\bm{F}_{\bm{\theta}} \, .
\end{align}
The final relation can be useful when thinking about Fisher information calculated in different physical units (different scalings). The only consequence of such a procedure is that the logarithm of the Fisher information shifts by a constant $C$, independently on the position in the parameter space. This further means that the distribution of the Fisher information shifts by the same constant.

For the single parameter variance definition (Eq. \ref{eq:single_parameter_variance}), the units (scaling) do not play a role, as we always express it with respect to the 2DPS.\ Therefore,\begin{equation}
    \frac{\sigma^2}{\sigma^2_{\rm 2DPS}} = \left(\frac{\det \bm{F}_{\bm{\theta}}}{\det \bm{F}_{\bm{\theta}, {\rm 2DPS}}}\right)^{-1/5} = \left(\frac{\det \bm{F}_{\bm{\eta}}}{\det \bm{F}_{\bm{\eta}, {\rm 2DPS}}}\right)^{-1/5} \, .
\end{equation}

{
 
\section{Convergence tests for the Fisher matrix} \label{app:convergence}
\begin{figure}
    \centering
    \includegraphics[width=0.95\linewidth]{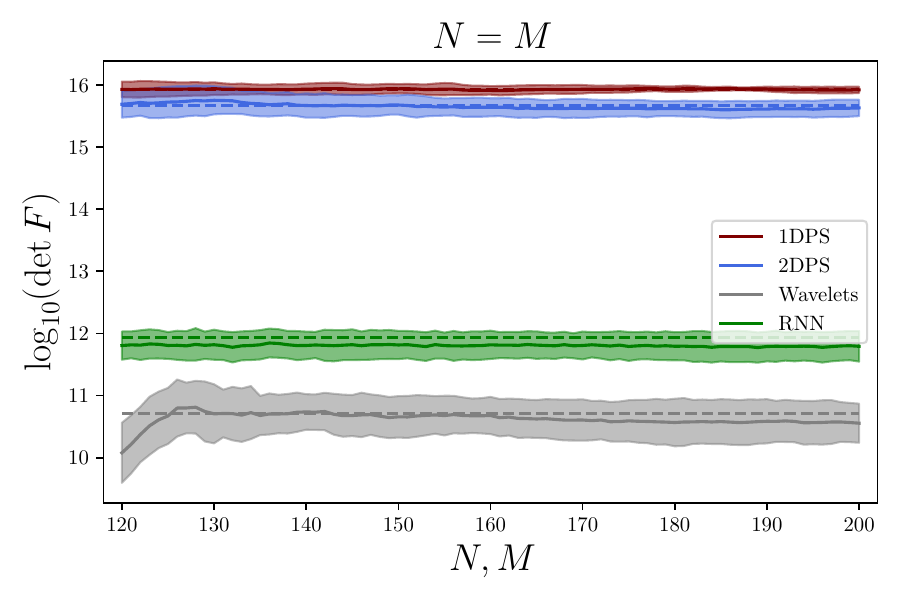}
    \caption{  Convergence test of the Fisher information at the fiducial parameter combination, $\bm{\theta}_{\text fid}$, as a function of $N = M$. The solid lines (shaded regions) correspond to the means (standard deviations) computed over five random realizations of summaries at $\bm{\theta}_{\text fid}$.  These can be compared to the dashed lines, which show the mean computed over the entire set of 1024 realizations. }
    \label{fig:app:convergence_NM}
\end{figure}
\begin{figure}
    \centering
    \includegraphics[width=0.95\linewidth]{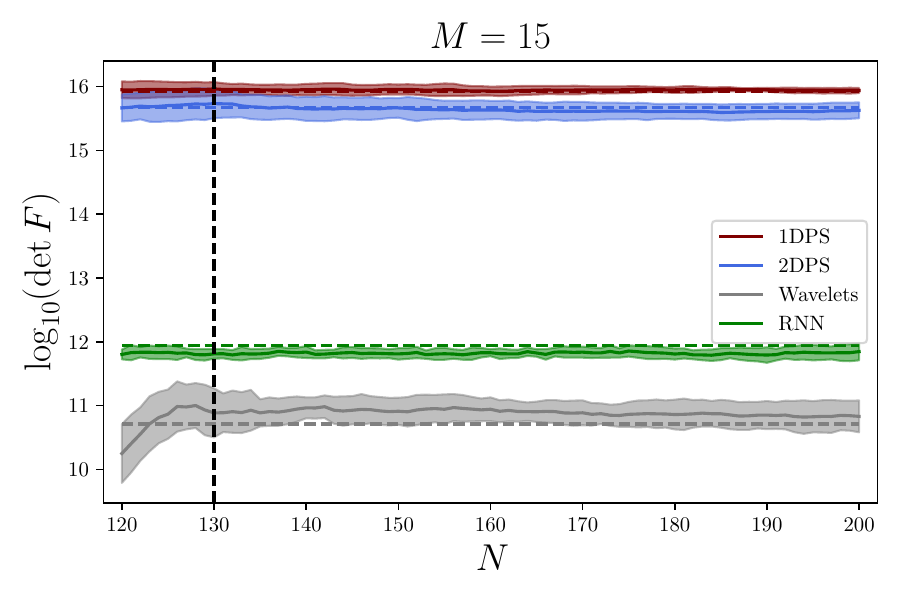}
    \caption{  Like Fig. \ref{fig:app:convergence_NM}, except here we fix $M = 15$ and vary $N$. The black vertical line denotes the fiducial value, $N = 130$, used in this work.}
    \label{fig:app:convergence_M_15}
\end{figure}
\begin{figure}
    \centering
    \includegraphics[width=0.95\linewidth]{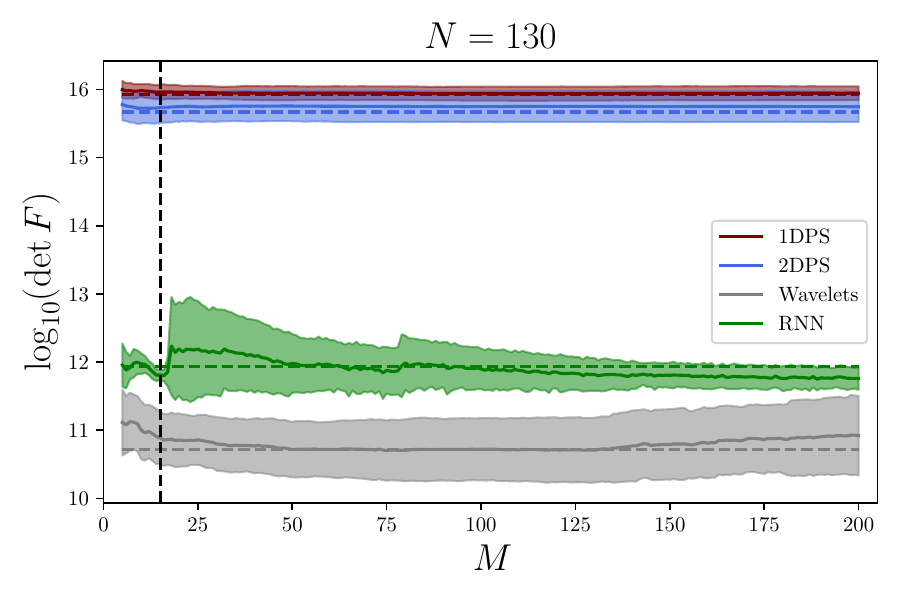}
    \caption{  Like Fig. \ref{fig:app:convergence_NM}, except here we fix $N= 130$ and vary $M$. The black vertical line denotes the fiducial value, $M = 15$, used in this work.}
    \label{fig:app:convergence_N_130}
\end{figure}
When computing the Fisher matrix (Eq. \ref{eq:Fisher_Gaussian}), one has to decide on the number of samples from which $\bm{\Sigma}^{-1}$, $\nabla_{\bm{\theta}}\bm{\mu}$ and $\nabla_{\bm{\theta}}\bm{\Sigma}$ are computed. In other words, one has to decide on the size of $N$, $M$ and $L$ (see Eqs. \ref{eq:mu_mean} - \ref{eq:sigma_der}). In this work, we fixed $M = L$ and used the same samples to compute $\nabla_{\bm{\theta}}\bm{\mu}$ and $\nabla_{\bm{\theta}}\bm{\Sigma}$. Furthermore, we fixed $N = 130$ and $M = 15$. We justified these choices by running a series of convergence tests.

For this test, we used a database of $1024$ simulations and their numerical derivatives centered at the fiducial parameters $\bm{\theta}_{\text fid}$, used to train the IMNN (see Section \ref{subsec:IMNN}). For that reason, we excluded the IMNN from this test -- as the summary space size for IMNN is only 5, it is the least problematic one considering the convergence.

In Figures \ref{fig:app:convergence_NM}, \ref{fig:app:convergence_M_15}, and \ref{fig:app:convergence_N_130}, we show three slices through the $N, M$ space. In Figure \ref{fig:app:convergence_NM}, we fix $N = M$, computing the Fisher information for each summary using five randomly sampled realizations.  The solid lines (shaded regions) correspond to the means (standard deviations) computed over these five subsamples.  These can be compared to the dashed lines, which show the mean computed over the entire set of 1024 realizations.
We see that all of our summaries have converged over the entire range shown except the wavelets, which converge after $N\geq130$.  This result is expected, as wavelets are $112$-dim summary, while others are significantly smaller.  

In Figure \ref{fig:app:convergence_M_15}, we fix $M = 15$ and vary $N$ in the same range as in Fig. \ref{fig:app:convergence_NM}. Here the black vertical line marks our fiducial choice of $N = 130$. {   We see that all summaries at $N = 130$ have converged, although the RNN and wavelets show evidence of a small residual bias. The later statistics require a larger $M$ to fully converge to the mean.}

Finally, in Figure \ref{fig:app:convergence_N_130} we fix $N = 130$ and vary $M \in [5, 200]$, marking $M = 15$ with a black vertical line. Here 1DPS and 2DPS stay almost constant, while RNNs and wavelets show some evolution. Non-monotonic evolution of variance in the RNN might be a consequence of small number of samples on which the lines have been computed. Most importantly, the chosen point of $M = 15$ shows a small bias for all summaries. The quality of the measured Fisher information might be better at $M \approx 100$; however, this would increase the database size by two orders of magnitude, which is computationally unreachable.  Importantly, these biases are over an order of magnitude smaller than the summary to summary and parameter to parameter variances computed in the main text; therefore our main conclusions are robust to our database size. {   However, we caution that RNN and wavelet summaries need a larger $M$ to fully converge.}

One could increase the numerical stability and statistical significance of the Fisher estimates by running a Bayesian inference over the covariance matrix and other parameters (see, e.g., \citealt{Alvarez2014}). This is noted in \cite{Hothi2023}, in which the authors study 21 cm Fisher forecasts using different wavelet-based summaries. In our setup, this would mean introducing a prior distribution over the covariance matrix, such as the inverse Wishart distribution \citep{Barnard2000}. One would then estimate the posterior distribution of the covariance from the samples of a summary and marginalize over it while recovering a distribution of the Fisher information. A small step in this direction has been made in this work by properly un-biasing estimates of the covariance and its inverse (Eqs. \ref{eq:sigma_unbiased} and \ref{eq:sigma_inv_unbiased}). A more complete statistical analysis is left for future work.
}
\end{appendix}

\end{document}